\documentclass[%
 reprint,
 amsmath,amssymb,
 aps,
pra,
]{revtex4-2}

\usepackage{graphicx}
\usepackage{dcolumn}
\usepackage{bm}

\usepackage{multirow}
\usepackage{lineno}
\usepackage[table]{xcolor}
\usepackage[normalem]{ulem}
\usepackage{color}

\newcommand{\rmin}{ {\rm in }}
\newcommand{\rmout}{ {\rm out }}
\newcommand{\rmscat}{ {\rm scat }}

\newcommand{\rmtot} { {\rm tot  }}
\newcommand{\BM  } { {\bm{{\cal M}}} }
\newcommand{\rminc}{ {\rm inc }}
\newcommand{\rmfar}{ {\rm far }}

\begin{document}

\preprint{APS}

\title{Controlled light scattering of a single nanoparticle by wavefront shaping}

\author{Peilong Hong$^{1,3}$}
\email{plhong@uestc.edu.cn}
\author{Willem L. Vos$^2$}
\email{w.l.vos@utwente.nl}
\affiliation{$^1$School of Optoelectronic Science and Engineering, University of Electronic Science and Technology of China (UESTC), Chengdu 611731, China \\
$^2$Complex Photonic Systems (COPS), MESA+ Institute for Nanotechnology, Faculty of Science and Technology, University of Twente, P.O. Box 217, 7500 AE Enschede, The Netherlands\\
$^3$The MOE Key Laboratory of Weak-Light Nonlinear Photonics, Nankai University, Tianjin 300457, China} 

\date{July 5th, 2022, in preparation for Phys. Rev. A}

\begin{abstract}
Controlling light scattering by nanoparticles is fundamentally important for the understanding and the control of light with photonic nanostructures, as well as for nanoparticle scattering itself, including Mie scattering. 
Here, we theoretically and numerically investigate the possibility to manipulate nanoparticle scattering {\color{black}by} wavefront shaping that was initially developed to control {\color{black} light scattered by large numbers of nanoparticles in nanophotonic media. }
{\color{black}By employing a scattering matrix analysis, we} find that even a single nanoparticle supports multiple strongly scattering eigenchannels, {\color{black}suggesting wavefront shaping as a promising tool to manipulate scattered light of a single nanoparticle.}
{\color{black}By sending in shaped wavefronts, we selectively excite eigenchannels, as is apparent from the distinct field distributions. }
{\color{black}These scattering eigenchannels are related to different resonant leaky modes of the scatterer, that reveal remarkable localized "hot spots" where the field is substantially enhanced. }
Moreover, we investigate {\color{black}the backscattered spectra; to this send in wavefronts relevant for a particular eigenchannel, and observe that the backscattered spectrum reveals not only the excited channel but also several others. 
This result points to the existence of short and long-range spectral correlations for an eigenchannel. }
Our work offers a {\color{black}flexible} tool to {\color{black}manipulate light scattering of a single nanoparticle, and thus opens new possibilities to} control {\color{black} field patterns and} light-matter interactions {\color{black} in a nanoparticle, as well as to explore new features of nanoparticle scattering such as the spectral correlation and temporal response of light scattered by} nano scatterers, including Mie spheres. 
\end{abstract}

\maketitle

\section{Introduction} \label{sec:Introduction} 
The manipulation of waves, such as electronic waves, acoustic waves, and optical waves, is an ongoing central topic in many research fields~\cite{Akkermans2007BOOK}. 
In optics, many different mesoscopic nanostructures have been developed to control light waves, including photonic crystals and scattering media~\cite{joannopoulos2008molding, Lourtioz2008Book, Ghulinyan2015Book}, plasmonic structures~\cite{Barnes2003Nature}, metamaterials~\cite{Soukoulis2011NatPhoton}, and metasurfaces~\cite{yu2014flat}. 
In particular, photonic scattering media that consist of many randomly-distributed nanoparticles, have been applied to control light propagation and light-matter interactions for, notably, Anderson localization and transverse localization~\cite{Lagendijk2009PhysToday, Wiersma2013NatPhot, Segev2013NatPhot, Skipetrov2016NJP, Skipetrov2020PRB}, weak localization and coherent backscattering~\cite{Vanalbada1985PRL, Wolf1985PRL, Fazio2017NatPhot}, random lasing~\cite{cao1999random, fallert2009co}, imaging through opaque media~\cite{bertolotti2012non, katz2014non, Hong2018APL, sarma2016control}, and sensing deep inside opaque media such as biological tissue~\cite{Durduran2010RepProgPhys, Mariani2018OE}. 
Recently, optical wavefront shaping (WFS)~\cite{vellekoop2007focusing, popoff2010measuring, mosk2012controlling} has been demonstrated to be a powerful method to manipulate light in and through complex scattering media, by spatially shaping the incident wavefront, leading to novel applications in high-resolution imaging~\cite{vellekoop2010exploiting, vanputten2011scattering, deaguiar2017polarization, yeminy2021guidestar, pai2021scattering}, enhanced energy delivery~\cite{choi2011transmission, Hong2018Optica, Uppu2021PRL}, efficient light emission~\cite{bachelard2012taming, qiao2017second, lib2020real}, and classical and quantum communication schemes~\cite{Goorden2014Optica, wolterink2016programmable, valencia2020unscrambling, leedumrongwatthanakun2020programmable}. 

Many classes of nanostructures such as photonic crystals, metasurfaces, and scattering media, consist of \textit{assemblies of many nanoscatterers}, hence light scattering plays a fundamental role in these nanostructures~\cite{kuznetsov2016optically,staude2017metamaterial}.
Consequently, manipulation of light scattering is crucial for the understanding and control of light propagation in these complex nanostructures. 
To control the scattering of a \textit{single nanoparticle}, typically the structural parameters are engineered, notably to tailor multipolar interference and scattering properties of the scatterer upon illumination by incident plane waves~\cite{geffrin2012magnetic, miroshnichenko2015nonradiating, shamkhi2019transverse}. 
When achieving resonant excitation for a nanoparticle, the excited cavity mode has been exploited to enhance light-matter and nonlinear interactions, such as,  photoluminescence~\cite{rutckaia2017quantum, cihan2018silicon},  lasing~\cite{gongora2017anapole,tiguntseva2020room}, and optical harmonic generation~\cite{grinblat2016enhanced, xu2018boosting}. 

{\color{black}A little explored question is whether the scattering of a single nanoparticle can also be manipulated by shaping the wavefronts of the incident light, in other words, by optical WFS. 
Indeed, first steps in this direction have recently been taken by Refs.~\cite{wei2016excitation, raybould2017exciting, saadabad2021structured} who proposed how to make non-radiating (anapole) resonances radiating by tweaking the near fields; the main difference to these seminal papers is that here we wish to shape wavefronts for nanoparticles by using far-field information. 
}

\begin{figure}[hbt] 
	\centering
	\includegraphics[width=0.48\textwidth]{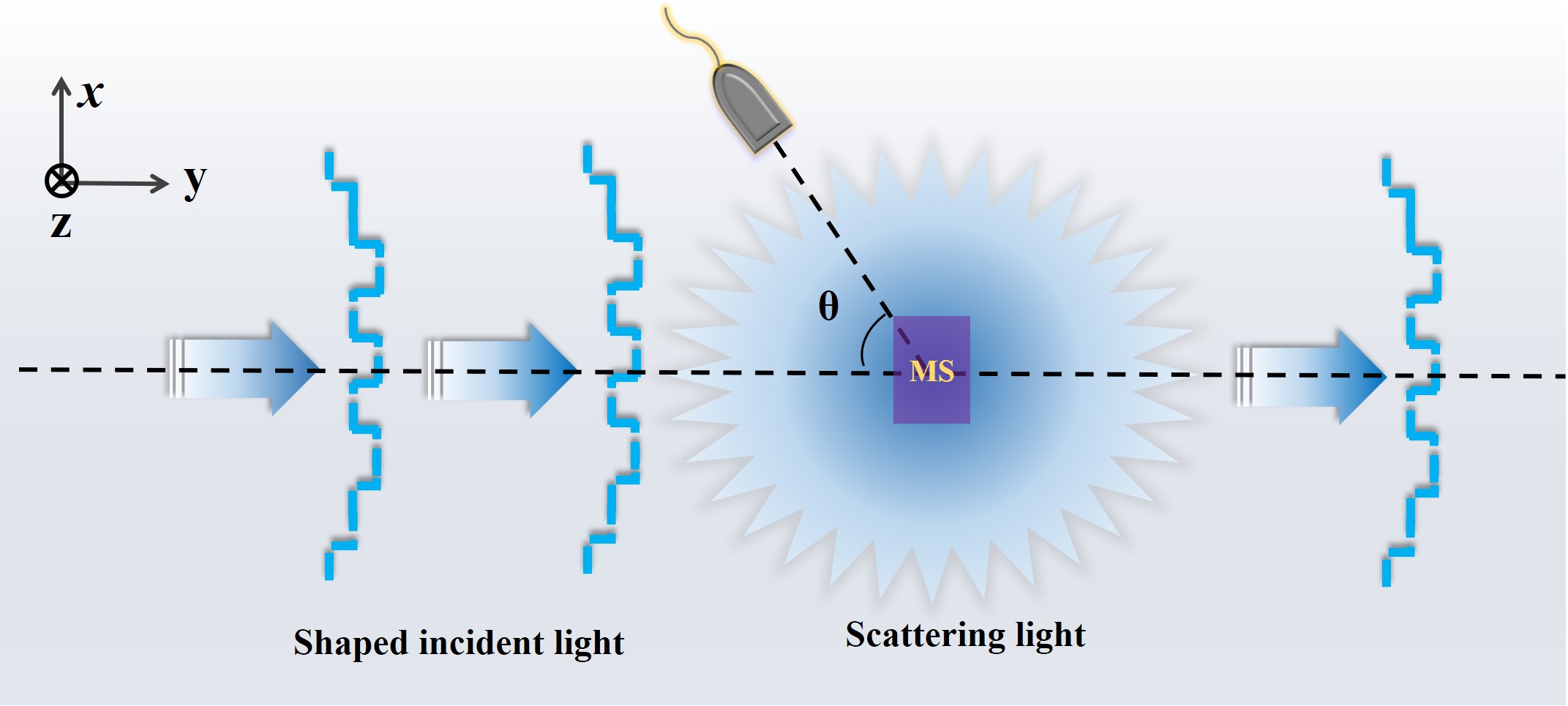}
	\caption{Schematic diagram of the physical situation studied here. 
	Incident light with a shaped wavefront is incident on a scattering particle (MS), which results in scattered waves in far field with controllable properties.
    This WFS diagram for manipulating Mie scattering can be realized practically, where the light scattered at an angle $\theta$ is monitored by a detector in the far field. 
   	\label{fig:Fig1_scheme}
   	}
\end{figure}

At first glance, the answer to the question above seems to be "no", since only a single optical path exists for light scattered at an angle $\theta$ by a \textit{single} nanoparticle, {\color{black}as illustrated in} Figure~\ref{fig:Fig1_scheme}.  
This situation fundamentally differs when the sample consists of \textit{multiple} ($N$) scatterers, in which case optical interference between multiple light paths can be controlled by WFS~\cite{vellekoop2007focusing, popoff2010measuring, Rotter2017RMP}. 
Remarkably, from our analysis in this paper of the {\color{black}{scattering}} matrix of a single scatterer that is experimentally accessible, we find that there are also \emph{multiple} strongly scattering eigenchannels for such a single scatterer. 
Therefore, optical WFS can effectively be employed to manipulate light scattering by selectively exciting these single-particle eigenchannels. 
The physical essence is the realization that light waves interacting with a single nanoparticle are not necessarily experiencing single scattering (in which case wavefront shaping control would not be feasible), but the waves are multiply scattered by the single nanoparticle, as described by the $t$-matrix~\footnote{The transition matrix, that originally originates in high-energy physics should not to be confused with the transmission matrix that originates from mesoscopic physics. }~of a single scatterer~\cite{Lagendijk1996PhysRep, vanRossum1999RMP, Rotter2017RMP}.
{\color{black}Nonetheless, such kind of multiple scattering in a single nanoparticle is essentially different from that in disordered media.
Our scattering eigenchannel analysis suggestes wavefront control as a promising tool to explore novel features of nanoparticle scattering, profiting from the flexible manipulation on the scattered light with WFS.}

By investigating the scattering light field inside the nanoparticle, we find that these highly scattering eigenchannels are related to different leaky resonances of the scatterer. 
Consequently, by employing optical WFS to selectively excite different scattering eigenchannels, it is possible to control resonance-enhanced light-matter interactions with individual nanoparticles. 
Moreover, we find that the highly scattering eigenchannels exhibit both short range and long range spectral correlations, that are related to the rich resonant features of the nanoparticle, and could lead to interesting ultrafast optics. 
These results offer new perspectives for understanding and controlling the intriguing light scattering process in various nanostructures. 


\section{Methods}\label{sec:methods} 
{\color{black}Figure~\ref{fig:Fig1_scheme} shows a schematic of shaped wavefronts that are incident on and scatter from a nanoparticle, where the wavefronts of the incident light are controlled by optical WFS equipment further upstream. }
The nanoparticle scatters light into {\color{black}all possible outgoing wave vectors, corresponding to all} angles in the far field{\color{black}, as described by the scattering matrix. }
{\color{black}Here, we focus on scattered waves that emanate in backscattered directions in the angular range ($-90^o < \theta < 90^o$), because these waves do not interfere with the incident waves that travel in the positive y-direction~\footnote{Thus our study is complementary to the recently discovered scattering phenomena of mutual extinction and transparency, where one explicitly considers the interference between scattered waves and the incident wave~\cite{lagendijk2020mutual}.}. }
To thoroughly analyze the backscattering, we employ the random matrix theory to describe the relationship of the incident and outgoing modes at different far-field angles~\cite{popoff2010measuring,vellekoop2008universal,yu2013measuring,goetschy2013filtering}, demonstrating the existence of multiple scattering eigenchannels with merely a single {\color{black}scattering nanoparticle} 
The existence of multiple highly scattering eigenchannels may shed new light on the recently reported scattering phenomena of mutual extinction and transparency~\cite{lagendijk2020mutual, rates2021observation} for various nanoparticles and nanostructures.

Following the random matrix theory~\cite{popoff2010measuring,vellekoop2008universal,yu2013measuring,goetschy2013filtering}, we model the incident and outgoing modes {\color{black} in free space} as column vectors, and obtain a scattering matrix ${\bm S_r}$ for the single nanoparticle. 
{\color{black}To be precise, the matrix ${\bm S_r}$ is a subset of the whole scattering matrix in view of our choice to consider the back scattered waves.} 
The incident and outgoing free {\color{black}space} modes are represented by the column vectors $\bm{L_{\rmin}}$ and $\bm{L_{\rmout}}$, respectively. 
The square of the $i$th element $|L^{\rmin}_i|^2$ or $|L^{\rmout}_i|^2$ denotes the {\color{black}optical} power in the relevant incident or scattering channel. 
Hence, the $i$th element is related to the electric field by $|L^{\rmin}_i|^2 = |E^{\rmin}_i|^2 \cdot \Delta S_{\rmin}$ (or  $|L^{\rmout}_i|^2 = |E^{\rmout}_i|^2 \cdot \Delta S_{\rmout}$ ), where $\Delta S_{\rmin}$ (or $\Delta S_{\rmout}$) is a constant value dependent on the area of the discrete incident (or scattering) channels.
Once the scattering matrix ${\bm S_r}$ is retrieved, {\color{black}in an experiment for instance by interferometry as illustrated in Appendix}, the backscattering by a single nanoparticle is obtained from the following expression 
\begin{equation}\label{eq:Sr_general}
     \bm{L_{\rmout}} = {\bm S_r} \bm{L_{\rmin}} \,.
\end{equation}
{\color{black}The main step in this study is that singular value decomposition on the matrix ${\bm S_r}$ will find the scattering eigenchannels, as given by }
\begin{equation}\label{eq:Sr_SVD}
\bm{S_r} = \bm{U} \bm{S} \bm{V}^+, 
\end{equation}
where $\bm{U}$ and $\bm{V}$ are unitary matrices, and $\bm{S}$ is a diagonal matrix~\cite{Press2007book}. 
{\color{black}Here, the diagonal elements $s_i$ of the matrix $\bm{S}$ are the eigenvalues of the scattering channels, and the columns of the unitary matrices $\bm{V}$ and $\bm{U}$ represent the eigenfunctions or wavefunctions of the incident and scattered channels, respectively~\cite{Miller2019AOP}. }
Therefore, the distribution of the diagonal elements $s_i$ holds information on the scattering features of the single scatterer. 
{\color{black}From the singular value decomposition, we find that a single scatterer supports many eigenchannels, of which a subset will appear to be strongly scattering. 
Next, we will show, both by theory and full wave calculations, that one can manipulate the scattered intensities of a single nanoparticle, since shaping the incident waves can selectively excite a scattering eigenchannel, or even controls the interference between waves scattered via different eigenchannels.}

{\color{black}The full-wave simulation is implemented in the frequency domain in Comsol Multiphysics for the 2D case.
In the computation domain, the nanoparticle is surrounded by an air layer.
A perfect matching layer is added at the boundary of the computation domain.
The discretization of the computation domain is set to be smaller than 1/12 wavelength.
The field pattern is computed up to a radius larger than the wavelength of the incident light, but are cropped to zoom in on the most interesting regions when it is shown in the following.}

\section{Scattering eigenchannels of a single scatterer}\label{sec:eigenchannels_Mie} 
\subsection{Nanoparticle in 2 dimensions}\label{sec:Eigenchannels_2D}
\noindent\textbf{Configuration.} We first consider a 2D nanoparticle with a rectangular shape, since this is straightforward to discretize on a rectangular computational grid. 
The height of the scatterer (along y axis) is a = 300 nm, and the width (along x axis) is b = 600 nm. 
The refractive index of the scatterer material is taken to be $n = 3.5$, typical of high-index semiconductors like Si or GaAs~\cite{green2008self,jellison1992optical}. 
The {\color{black}{scattering}} matrix ${\bm S_r}$ is obtained by sequentially sending plane incident waves at different incident angles and simultaneously monitoring the scattered field in the far field plane. 
In our full wave simulation, a $z$-polarized incident light $E_{\rminc}(\theta_\rmin)$ is scanned from -45$^o$ to 45$^o$ (anticlockwise from $\hat{y}$) with a stepsize $\Delta\theta_\rmin$ (= 1$^o$).
This scanning range corresponds to a numerical aperture $NA = 1/\sqrt{2} \approx 0.71$, which is practically realizable with a commercial objective lens. 
Meanwhile, the backscattered field $E_{\rmfar}(\theta_\rmout)$ with $\theta_\rmout \in$ [-45$^o$, 45$^o$] (anticlockwise from -$\hat{y}$) is obtained at a sampling step $\Delta\theta_\rmout$ (= 1$^o$), which is easily measurable in practice with the same objective lens used for sending the incident light.

\vspace{12pt}\noindent\textbf{Methods.} With the above configuration, the incident and backscattered electric field is described by a column vector $\bm{E}_{\rmin}$ ($\bm{E}_{\rmout}$) with $N (=91)$ elements, each element of which is the electric field at a specific angle. 
The backscattered light field is related to the incident light field by employing a scattering matrix $\bm{M}$ with dimensions $N\times N$ (here N = $91$), expressed as 
\begin{equation} \label{eq:M_2D}
\bm{E}_{\rmout} = \bm{M} \bm{E}_{\rmin}. 
\end{equation}
By sequentially scanning the angle of incident light and monitoring the backscattered light, we obtain the scattering matrix $\bm{M}$.
To further relate $\bm{M}$ to power-related {\color{black}{scattering}} matrix $\bm{S_r}$, we need to calculate the power of each incident channel against the light field, as well as that of the {\color{black}{scattering}} channels.
For $i$th incident channels, the power of light illuminating on the scatterer is expressed as
\begin{equation}\label{eq:Pin_2D}
P^{\rmin}_i \equiv |L^{\rmin}_i|^2  = |E^{\rm{in}}_i|^2 \cdot \sigma_{\rmscat}/\Delta\theta_\rmin \,,
\end{equation} 
where $E^{\rminc}_i$ is the electric field of the incident plane wave from the $i$th-channel, and
$\sigma _{\rmscat}$ denotes the cross size of the scatterer multiplied by a unit length of $z$ axis.
For a subwavelength scatterer, the value of $\sigma _{\rmscat}$ could be set as the length of Airy spot $ \lambda/n_1$ in 2D. Here, $\lambda$ is the wavelength, and $n_1$ is the refractive index of the material surrounding the scatterer. 
Then, we derive to obtain the column vector $\bm{L_{\rmin}}$ as
\begin{equation} \label{eq:Lin_2D}
\bm{L_{\rmin}} = \bm{E_{\rmin}} \cdot \sqrt{\sigma_{\rmscat}/\Delta\theta_\rmin} \,.
\end{equation}
The vector $\bm{L_{\rmin}}$ is related to the total input power $P_{\rmin}^{\rmtot}$ as $P_{\rmin}^{\rmtot} = \bm{L_{\rmin}}^+ \bm{L_{\rmin}}$.
By considering the cylindrical surface of the far-field scattering light, the power $P_\rmout$ of scattered light is expressed as 
\begin{equation}\label{eq:Pout_2D}
P^{\rmout}_j \equiv |L^{\rmout}_j|^2 = |E^{\rmfar}_j|^2 \cdot {\color{black} r} \Delta\theta_\rmout\,,
\end{equation}
where {\color{black}$r$} is radius of the far-field circle.
Then, we derive to obtain the column vector $\bm{L_{\rmout}}$ as
\begin{equation} \label{eq:Lout_2D}
\bm{L_{\rmout}} = \bm{E}_{\rmout} \cdot \sqrt{{\color{black} r} \Delta\theta_\rmout } \,.
\end{equation}
The matrix $\bm{L_{\rmout}}$ is related to the total output power $P_{\rmout}^{tot}$ as $P_{\rmout}^{tot} = \bm{L_{\rmout}}^+ \bm{L_{\rmout}}$. 
Now, by substituting Eqs.~\ref{eq:Lin_2D} and \ref{eq:Lout_2D} to Eq.~\ref{eq:Sr_general},
the {\color{black}{scattering}} matrix $\bm{S_r}$ is derived to be
\begin{equation}\label{eq:Sr_2D}
\bm{S_r}= \beta_{2D} \cdot \BM \,
\end{equation}
where the constant prefactor is equal to $\beta_{2D} =  \sqrt{{\color{black} r}\Delta\theta_\rmout \Delta\theta_\rmin /\sigma_{scat}}$.

\vspace{12pt}\noindent\textbf{Results.} With the {\color{black}{scattering}} matrix $\bm{S_r}$, we carried out the singular value decomposition by using Eq.~\ref{eq:Sr_SVD}.,
We find that the nanoparticle supports $N$ eigenchannels, of which a subset $N_{s}$ appears to be strongly scattering. 
To quantify strongly scattering, we propose as a criterion to relate eigenvalues of higher order channels to the first, strongest, one $(s_j / s_1)$ and take as a gauge that the $j$th channel exceeds $(s_j / s_1 > 10 \%)$ compared to the strongest one. 

\begin{table}[htb]
	\centering
\setlength{\tabcolsep}{1mm}{
	\begin{tabular}{ccccccccc}
\\
\hline
	$b/\lambda$ & 1/4  & 2/7  & 1/3    & 2/5   & 1/2   & 3/5   & 3/4   & 1     \\ 
\hline
		\multicolumn{1}{c|}{\cellcolor[HTML]{FFFFFF}$s_1$} & 0.520 & 0.527 & 0.586 & 0.686 & 0.667 & 0.267 & 0.603 & 0.830 \\
		\multicolumn{1}{c|}{\cellcolor[HTML]{FFFFFF}$s_2$} & 0.158 & 0.129 & 0.135 & 0.350 & 0.356 & 0.216 & 0.411 & 0.414 \\
		\multicolumn{1}{c|}{\cellcolor[HTML]{FFFFFF}$s_3$} & 0.001 & 0.003 & 0.015 & 0.012 & 0.033 & 0.060 & 0.371 & 0.298  \\
\hline
	\end{tabular}}
	\caption{The eigenvalues of the first three highest scattering eigenchannels of the {\color{black}{scattering}} matrix at different frequencies. 
	The results show that the subwavelength scatterer has $N_{s} > 1$ strongly scattering eigenchannels.}
\label{table:Tab1}
\end{table}

In Table~\ref{table:Tab1}, we list the first three eigenchannels with substantial eigenvalues of the {\color{black}{scattering}} matrix with increasing frequency.
Here, we consider the wavelength range with the incident wavelength larger than the scatterer's dimension $b$. 
At the lowest frequencies $b/\lambda =1/4, 2/7, 1/3, 2/5, 1/2$, the ($j=2$) eigenvalue is strong, and the ($j = 3$) eigenvalue is weakly scattering. 
Thus, the number of strongly scattering channels is $N_s = 2$. 
At the higher frequencies $b/\lambda =3/5, 3/4, 1$, both the ($j=2$) and ($j = 3$) eigenvalues are strongly scattering, thus $N_s = 3$. 
It is thus clear that the number of eigenchannels is greater than one ($N_{s} > 1$, as naively expected in the introduction), in other words, more than one highly scattering eigenchannels is sustained by the nanoparticle. 
By further increasing the size of the scatterer (or equivalently, increasing the frequency of incident light), more highly scattering channels are expected to appear. 
These results demonstrate that manipulation of nanoparticle scattering is feasible with optical WFS, by selectively addressing these different strongly scattering eigenchannels.

\subsection{Mie sphere in 3 dimensions}\label{sec:Eigenchannels_3D}
\noindent\textbf{Configuration.} Generally, a full-wave simulation for an arbitrary {\color{black}three-dimensional} (3D) nanoparticle, including a Mie sphere, will require computationally a more intensive scan of the incident beam than in 2D, which is thus much more time consuming.
Therefore, we focused on the Mie sphere that has analytical solutions on the scattering light~\cite{Bohren2008Book,Frezza2018JOSAA}.
Without losing generality, we did theoretical calculations for a Mie sphere with diameter $d_s =$ 400 nm and refractive index $n = 2$, typical for a high-index material such as silicon nitride. 
The normalized frequency of incident light is taken to be $d_s/\lambda = 8/9$. 
Again, we consider the backscattering configuration, in which the incident angle of light $\theta_{\rmin} \in [0^o,32^o]$ (relative to $\hat{y}$), and $\phi_\rmin \in [0^o,360^o) $ (relative to $\hat{x}$ in the $xz$ plane).
For the backscattered light, we have $\theta_{\rmout} \in [148^o, 180^o]$, and $\phi_\rmout \in [0^o,360^o) $. 
This setting can be realized with a microscopic setup with numerical aperture NA = $\sin(32^o) = 0.53$.
Subsequently, the electric field of the backscattered light ${\bm E}_{\rmfar}(\theta_\rmout, \phi_\rmout)$ is calculated analytically in a far-field spherical surface of large radius $R$ ($=1$ m), with an incident light ${\bm E}_{\rminc}(\theta_\rmin,\phi_\rmin)$. 
To obtain the {\color{black}{scattering}} matrix $\bm{S_r}$, we sample $\theta_\rmin$ with $N$ steps, and $\phi_\rmin$ with $M$ steps, and consequently the size of $\bm{E_{\rmin}}$ is $N_{\rmin} = 2NM \times 1$, where the factor 2 is introduced due to the two independent polarization of the light field.
For simplicity, we sample the backscattered light in a similar way as the incident field, and the size of $\bm{E_{\rmout}}$ is also $N_{\rmout} = 2NM \times 1$.
Here, the sampling steps $N$ is 8, and $M$ is 72. 
Therefore, the size of $\bm{L_{\rmin}}$ and $\bm{L_{\rmout}}$ is 1152, and the size of $\bm{S_r}$ is $1152\times 1152$.

\vspace{12pt}\noindent\textbf{Methods.} 
By continuously scanning the angle of incident light and monitoring the backscattered light, we get a scattering matrix $\bm{M}$ that relates the scattering field vector $\bm{E}_{\rmout}$ to the incident field vector $\bm{E}_{\rmin}$, \textit{i.e.}, 
\begin{equation} 
\label{eq:M}
\bm{E}_{\rmout} = \BM \, \bm{E}_{\rmin} \,,  
\end{equation}
Similarly as in the 2D case, we calculate the power-related {\color{black}{scattering}} matrix $\bm{S_r}$. 
The power of $i$th input channel that illuminates on the scatterer is given by 
\begin{equation}\label{eq:Pin_3D}
  P^{\rmin}_i \equiv   |L^{\rmin}_i|^2 = |\bm{E}^{\rminc}_i|^2 \cdot \sigma_{\rmscat}/(\Delta\theta_\rmin \Delta\phi_\rmin) \,,
\end{equation}
where $\sigma _{\rmscat} $ denotes the cross section of the Mie sphere, and could be set as the the area of airy spot $\pi (0.61\lambda/n_1)^2$ for a subwavelength scatterer. Here $\lambda$ is the wavelength, and $n_1$ is the refractive index of the environment material surrounding the scatterer. 
Then, we obtain the column vector $\bm{L_{\rmin}}$ as
\begin{equation} \label{eq:Lin_3D}
  \bm{L_{\rmin}} \equiv \bm{E_{\rmin}} \sqrt{\sigma_{\rmscat}/(\Delta\theta_\rmin \Delta\phi_\rmin)} \,.
\end{equation}
The vector $\bm{L_{\rmin}}$ is related to the total input power $P_{\rmin}^{\rmtot}$ as $P_{\rmin}^{\rmtot} = \bm{L_{\rmin}}^+ \bm{L_{\rmin}}$. 
For the scattered light, the power of $j$th scattering channel at the far-field sphere surface is expressed as
\begin{equation}\label{eq:Pout_3D}
  P^{\rmout}_j = |\bm{E}^{\rmout}_j|^2 \cdot {\color{black} r}^2 \sin(\theta^\rmout_j) \Delta\theta_\rmout \Delta\phi_\rmout \,,
\end{equation}
We obtain the column vector $\bm{L_{\rmout}}$ as
\begin{equation} \label{eq:Lout_3D}
  \bm{L_{\rmout}} = \bm{G}(\theta^\rmout_j)  \bm{E}_{\rmout} \cdot {\color{black} r}\sqrt{\Delta\theta_\rmout \Delta\phi_\rmout } \,.
\end{equation}
Here, $\bm{G}(\theta_j^\rmout)$ is a $N_{\rmout}\times N_{\rmout}$ matrix with the value of elements at $\theta^\rmout_j$ being $\sqrt{\sin(\theta_j^\rmout)}$. 
The matrix $\bm{L_{\rmout}}$ is related to the total output power $P_{\rmout}^{tot}$ as $P_{\rmout}^{tot} = \bm{L_{\rmout}}^+ \bm{L_{\rmout}}$. 
Now, by substituting Eqs.~\ref{eq:Lin_3D} and \ref{eq:Lout_3D} to Eq.~\ref{eq:Sr_general},
the backscattering matrix $\bm{S_r}$ is derived to be
\begin{equation}\label{eq:Sr}
  \bm{S_r}= \beta_{3D} \cdot \bm{G}(\theta_\rmout) \BM \,
\end{equation}
where the constant prefactor is equal to $\beta_{3D} = {\color{black} r} \sqrt{\Delta\theta_\rmout \Delta\phi_\rmout \Delta\theta_\rmin \Delta\phi_\rmin /\sigma_{scat}}$.

\begin{table}[htb]
	\centering
\setlength{\tabcolsep}{1mm
\begin{tabular}{ccccccc}
\\
\hline
\multicolumn{1}{c|} {eigenchannel (j)} & 1   & 2    & 3   & 4   & 5   & 6 \\
\hline
\multicolumn{1}{c|} {$s_j$}   & 0.673  & 0.673  & 0.270  & 0.256  & 0.186  & 0.186 \\
\hline
\end{tabular}}
\caption{The eigenvalues of the first six highest scattering         eigenchannels for a subwavelength Mie sphere.
	The normalized diameter of the Mie sphere is $d_s/\lambda = 8/9$, and refractive index of the scatterer is 2.}
\label{table:Tab2}
\end{table}

\vspace{12pt}\noindent\textbf{Results.} By employing singular value decomposition for the {\color{black}{scattering}} matrix $\bm{S_r}$, the scattering eigenchannels of the Mie sphere are obtained, of which the first six highest eigenvalues are listed in Table~\ref{table:Tab2}. 
Using the same criterion for strongly or weakly scattering channels as above, it is clear that this subwavelength Mie sphere supports no less than ($N_s = 6$) strongly scattering eigenchannels.

\section{Field patterns for select eigenchannels }\label{sec:resonances_Mie} 
{\color{black}To obtain more insights on the scattering eigenchannels, we now investigate the field pattern inside and around the nanoparticle while selectively exciting an eigenchannel.
This is accomplished by effectively performing optical WFS in the full-wave simulation; to this end, we set the wave front of the incident wave by employing the structured eigen-wavefront from the matrix $\bm{V}$, and then compute the scattered field pattern at the near zone of the nanoparticle through full-wave simulation.}
For simplicity and easy of plotting, we focus on the 2D scatterer and the eigenvalues shown in Table~\ref{table:Tab1}.

\begin{figure}[htb]
	\centering
	\includegraphics[width=0.45\textwidth]{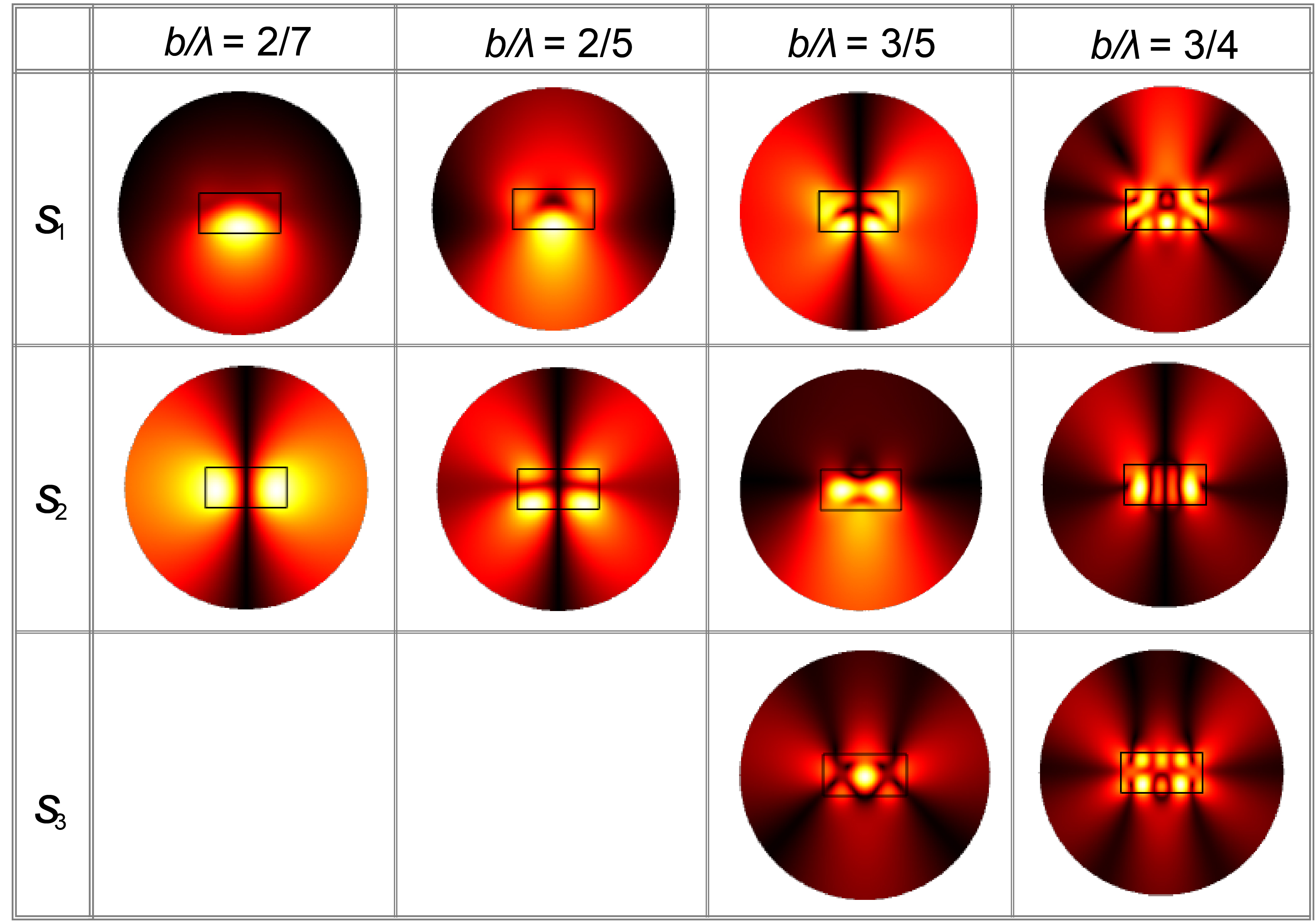}
	\caption{{\color{black}Field patterns inside and about the nanoparticle for individually excited eigenchannels at different frequencies. 
	The shaped waves are incident from the top to the bottom, and the field shown is only the scattered field.}
	From the first column to the fourth column, the frequencies of the incident light are $b/\lambda$ = 2/7, 2/5, 3/5 and 3/4, respectively.
	{\color{black}For clear visualization}, $|E_{scat}|$ in each panel has been normalized, such that the change of $|E_{scat}|$ from maximum to minimum is represented by the thermal colors changing from black to white. 
	The results show that different resonant leaky modes of the scatterer are excited. \label{fig:Fig2_ScatField}}
\end{figure}

Figure~\ref{fig:Fig2_ScatField} shows the scattered fields (without the incident field) inside and nearby the nanoparticle for different eigenchannels at four different frequencies $b/\lambda$ = 2/7, 2/5, 3/5 and 3/4, respectively.
{\color{black}At the lowest frequency $b/\lambda$ = 2/7, the first wave front excites a scattered field that has a maxima near the backside of the nanoparticle, without lobes, somewhat like a monopole. 
The second wave front excites a scattered field with a nodal plane through the axis of the nanoparticle, and two maxima on the left and right surfaces of the nanoparticle, akin to a dipole pattern (in 2D). 
At frequency $b/\lambda$ = 2/5, the first wave front excites a scattered field with a nodal line near the center of the nanoparticle and a maxima near the backside of the nanoparticle.
The second wave front excites a scattered field with two nodal planes perpendicular to each other and four maxima are located at the four corners of the nanoparticle, respectively.
At frequency $b/\lambda$ = 3/5,
the first wave front excites a scattered field with a curved nodal strip at the center of the nanoparticle and two maxima separated by a nodal plane along the axis.
The second wave front excites a scattered field with a pair of connected maxima at the center of the nanoparticle.
The third wave front excites a complex scattered field with a maxima located at the center of the nanoparticle, and five faint hot spots encircling the maxima.
At frequency $b/\lambda$ = 3/4,
the first wave front excites a complex scattered field, of which a pair of tilting hot strips locate near the top surface, and three hot spots locate near the bottom surface of the nanoparticle.
The second wave front excites a scattered field with a nodal line along the axis of the nanoparticle and four hot spots located symmetrically with respect to the nodal line.
The third wave front excites a complex scattered field with five hot spots near the top surface and other five hot spots near the bottom surface, but the magnitudes of the hot spots are not homogeneous.}
{\color{black}Generally,} more high-order resonant modes are excited {\color{black}as the frequency increases}, featured by more complex field patterns inside the scatterer.
{\color{black}These results show the relationship between the scattering eigenchannels and the resonant leaky modes inside and around the nanoparticle.}
From these results, we see that the excitation of different highly scattering eigenchannels selectively couples light into different resonant leaky modes of the nanoparticle.
{\color{black}In other words}, for optical scattering of a single nanoparticle, the existence of more than one highly scattering eigenchannel can be attributed to these resonant leaky modes.

From the full-wave simulation results, we also obtained {\color{black}the scattered light at the far field.
Then, we calculated the} ratios of the backscattered power to that of incident power, expressed as $R_j$. 
The simulation results fully agree with the prediction of the scattering matrix theory, given by 
\begin{equation}\label{eq:Sr_SVD_R}
R_j(\omega) = {s_j}^2 \,, 
\end{equation}
where $s_j$ denote the eigenvalues listed in Table~\ref{table:Tab1}. 

\section{Spectra of the scattering eigenchannels} \label{sec:Spectral_correlation} 

{\color{black}To reveal the resonant nature of the scattering eigenchannels, we explore the spectra of individual eigenchannels by scanning the frequency of incident light.}
Notably, the phase structure on the wavefront is mostly important for optical WFS, and pure-phase control has been applied in many WFS experiments~\cite{yeminy2021guidestar,pai2021scattering,Uppu2021PRL,valencia2020unscrambling, leedumrongwatthanakun2020programmable,Aulbach2011PRL, Katz2011NatPhot,McCabe2011NatComm,Mounaix2016PRL}. 
Following these considerations, we sent purely phase structured light to the scatterer, with the phase structures obtained from the first two strongly scattering eigenchannels at frequency $b/\lambda$ = 3/5. 
Subsequently, we scanned the frequency in the full wave simulation to obtain the scattering light field in the far field. 
In the end, we calculated to obtain the ratios of the backscattered power to that of incident power {\color{black}$R_j$} at different {\color{black}frequencies for the $j$th eigenchannel}.

\begin{figure}[htb]
	\centering
	\includegraphics[width=0.5\textwidth]{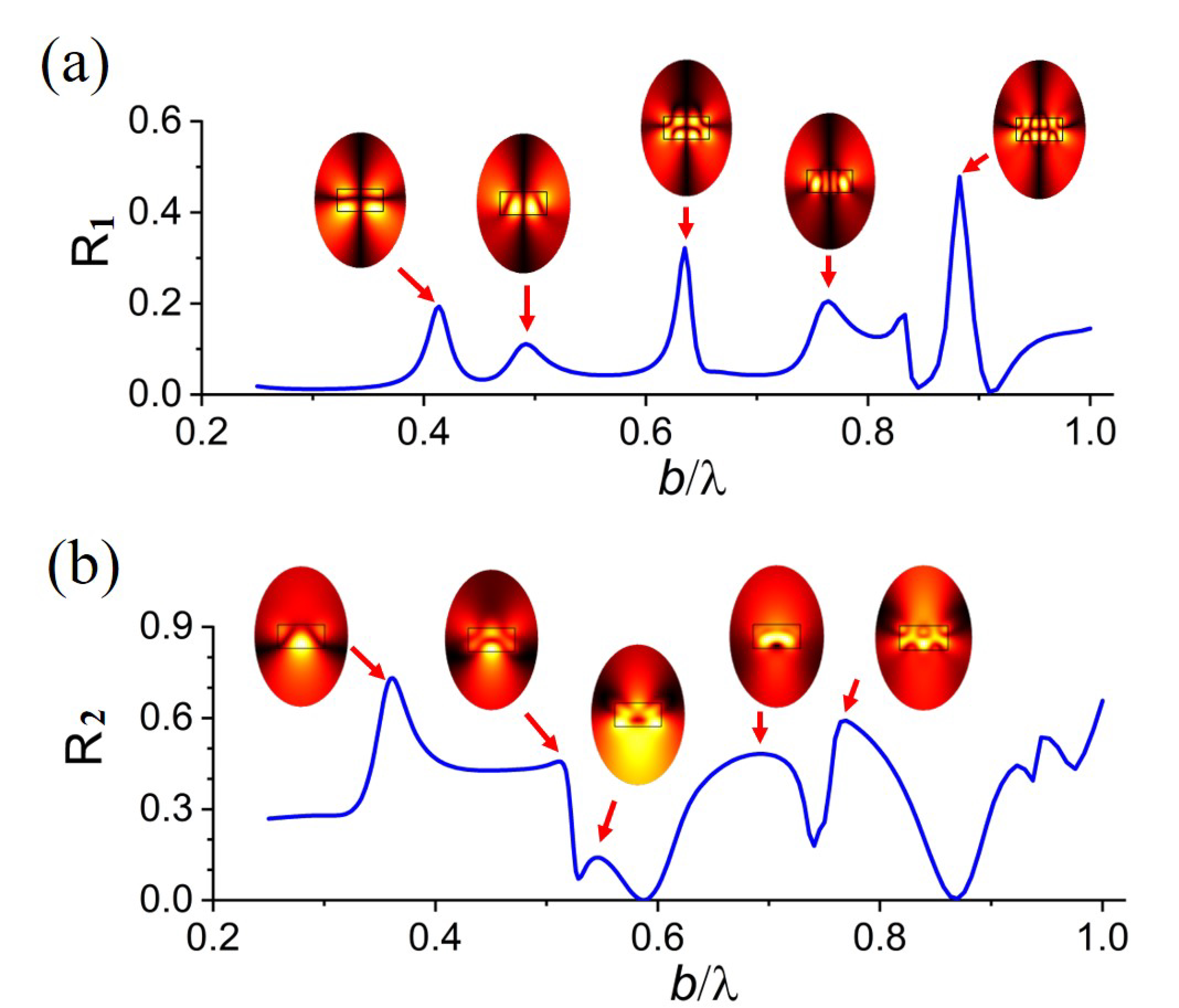}
	\caption{Relative backscattered power {\color{black}$R_j$ ($j=1,2$)} versus the frequency of the incident light with different eigen-wavefronts. 
	(a) The incident wavefront is loaded with the phase structure of the first eigenchannel ($s_1$=0.267) at frequency $b/\lambda$ = 3/5. 
	(b) The incident wavefront is loaded with the phase structure of the second eigenchannel ($s_2$ = 0.216) at frequency $b/\lambda$ = 3/5. 
	The results show that multiple peaks of backscattered power emerge in different spectral ranges relative to the initial frequency $b/\lambda$ = 3/5, indicating the coexistence of short range and long range spectral correlations for the highly scattering eigenchannels.
	\label{fig:Fig3_Correlation}}
\end{figure}

The results of the spectra for the two eigen-wavefronts are shown in Figs.~\ref{fig:Fig3_Correlation}(a) and~\ref{fig:Fig3_Correlation}(b), respectively.
The {\color{black}spectra} exhibit complex patterns {\color{black}constituted by resonant peaks at both the short range and the long range}.
At frequencies near the initial frequency $b/\lambda$ = 3/5, {\color{black} a single peak appears in the spectra}, as indicated by the right third and second red arrows in Figs.~\ref{fig:Fig3_Correlation}(a) and~\ref{fig:Fig3_Correlation}(b), respectively.
We have calculated the field patterns at the short-range {\color{black}resonant peak}, as shown in the insets of Figs.~\ref{fig:Fig3_Correlation}(a) and~\ref{fig:Fig3_Correlation}(b). 
{\color{black}The obtained field patterns are similar as that shown in Fig.~\ref{fig:Fig2_ScatField} at frequency $b/\lambda$ = 3/5.}
{\color{black}Such a single peak indicates the short-range spectral correlation of the individual eigenchannels, originating from an individual resonant leaky mode residing within the limited spectral range.}
For frequencies far away from the initial frequency $b/\lambda$ = 3/5, multiple {\color{black}resonant} peaks emerge. We have also calculated the field patterns at these long-range {\color{black} resonant} peaks shown in the insets of Figs.~\ref{fig:Fig3_Correlation}(a) and~\ref{fig:Fig3_Correlation}(b).
{\color{black}The field patterns inside the nanoparticle are significantly different between these long-range resonant peaks.
Nonetheless, these resonant peaks are excited by the same eigen wavefront of the scattering matrix, indicating the existence of the long-range spectral correlation for an eigenchannel.}
Besides,
there is an even number of hot spots horizontally for {\color{black}the resonant modes} in Fig.~\ref{fig:Fig3_Correlation}(a), while the number of hot spots is odd horizontally for {\color{black}the resonant modes} in Fig.~\ref{fig:Fig3_Correlation}(b).
It indicates that the long range correlation may originate from a series of resonant leaky modes that {\color{black}have} similar spatial features.

{\color{black}We note that analogous spectral correlations have been explored for light that is multiply scattered in randomly scattering media~\cite{pena2014single,shi2015dynamic,bosch2016frequency}, but not for a single nanoparticle. 
These intriguing spectra of the scattering eigenchannels constitute a first step in exploring novel spectral correlations of the light scattered by individual nanoparticles.}

\section{Perspective on applications} \label{sec:applications} 
{\color{black}In this section, we present as a brief outlook two different classes of possible applications that could be explored based on our results presented here. }

\subsection{Light-matter interactions}\label{sec:light-matter}
{\color{black} By transforming the regime of single incident plane wave to that of multiple incident plane waves, we can fully address the multiple strongly scattering eigenchannels intrinsic to a single nanoparticle.
Wavefront control on the scattering eigenchannels not only manipulate the scattered light in the far field, but also couples light into different resonant leaky modes sustained by the nanoparticle as shown in Fig.~\ref{fig:Fig2_ScatField}. 
By placing a reporting fluorophore at a select position in or near the scattering nanoparticle, by detecting the fluorescence in response to the tuned wavefront one can characterize the field strength within the nanoparticle.
Besides, wavefront control on the phase difference between two scattering eigenchannels allows one to control the local field strength, that can be reported by the fluorophore. 

The option described above is also useful to actively tune light-matter interactions on the nanoscale with a nanoparticle, such as fluorescence, lasing, non-linear optics including harmonic generation, and so on.
For instance, the distributions of field maxima inside the nanoparticle differ for different resonant modes. 
By shaping the incident wavefront, one could thus selectively excite select hot spots within and around the nanoparticle, and thereby activate light-matter interaction locally such as fluorescence of different kinds of two-level quantum systems (e.g., atoms, molecules, ions, quantum dots) implanted at different locations within the nanoparticle~\cite{Barnes2020JO}. 
}

\subsection{Ultrafast time-dependent physics}\label{sec:time-dependent}

{\color{black}The temporal response of a nanoparticle is related (by Fourier arguments) to the spectral properties of the strongly scattering channels, as discussed in section~\ref{sec:Spectral_correlation}. }
To investigate the temporal response of the scattering light, we focus on the 2D case for simplicity.
{\color{black} Let us consider time-dependent scattered light, where} the total scattering power within the detection range is proportional to 
\begin{equation}
\begin{split}
       P(t) &   =  \langle |E_{\rmfar}(t, \theta_\rmout)|^2 \rangle_{\theta_\rmout} \cdot {\color{black} r} \theta_\rmtot \\
        =  &  \mathcal{F}\Big\{ {\color{black} r} \theta_\rmtot \int \text{d}\omega  
          \langle  E^{*}_{\rmfar}(\omega, \theta_\rmout)E_{\rmfar}(\omega+\Delta \omega, \theta_\rmout)\rangle_{\theta_\rmout}   \Big\} \,.
\end{split}
\end{equation}
Here, $\langle \cdots \rangle_{\theta_\rmout}$ denotes the average over ${\theta_\rmout}$, {\color{black}$\theta_\rmtot$ means the detected angle range, and $r$ is the radius of the circular plane at the far field.}
It is seen that the temporal response of the nanoparticle is determined by the {\color{black}correlation function}
\begin{equation}
    C(\omega,\Delta \omega) = {\color{black} r} \theta_\rmtot \langle E^{*}_{\rmfar}(\omega, \theta_\rmout)E_{\rmfar}(\omega+\Delta \omega, \theta_\rmout)\rangle_{\theta_\rmout} \,.
\end{equation}
Since the overall scattered light can be decomposed to the scattering eigenchannels (column of $\bm{U}$), the {\color{black}correlation function} can be further computed over these eigenchannels.
Since the different eigenchannels are orthogonal, $C(\omega,\Delta \omega)$ can be derived to be
\begin{equation}\label{eq:C}
    C(\omega,\Delta \omega) = \sum_{i=1}^N \alpha^{*}_i(\omega) \alpha_i(\omega+ \Delta \omega)\cdot s_i(\omega) s_i(\omega + \Delta \omega)  \,,
\end{equation}
where $\alpha_i(\omega)$ is the coefficient of the incident light coupling into the $i$-th eigenchannel.
Clearly, only the strongly scattering {\color{black}eigenchannels} make significant contributions to the {\color{black}correlation function}.

In other words, from {\color{black}frequency-dependent backscattered powers $R_j$ as shown in Fig.~\ref{fig:Fig3_Correlation}}, one gets the factors $s_i(\omega) s_i(\omega + \Delta \omega)$ in Eq.~\ref{eq:C} to predict the temporal response of a single scatterer, and thus the response to short optical pulses. 
This is a single-particle analogy to time-resolved optical WFS on samples with many nanoparticles, see, \textit{e.g.}, Refs.~\cite{Aulbach2011PRL, Katz2011NatPhot, McCabe2011NatComm, Mounaix2016PRL}. 

{\color{black}For example, let us consider a simple scenario that a 10-fs laser pulse centred at 800 nm is incident on the nanoparticle.
Let us assume that the center frequency of the fs laser is at $\sim b/\lambda = $0.88, and hence the spectral range of the fs laser ( $\sim 27$\% of the center frequency) covers a significant range of the resonant peak shown in Fig.~\ref{fig:Fig3_Correlation}(a).
With the incident light coupled into the first eigenchannel, the light pulse is strongly scattered by the nanoparticle, especially at the center frequency.
In contrast, by coupling incident light into the second eigenchannel, the component at the center frequency is weakly scattered, as indicated by the spectral trough in Fig.~\ref{fig:Fig3_Correlation}(b).
Clearly, the temporal responses of the nanoparticle in the two situations are quite different, affected by the particular eigenchannel that is selectively excited. 
}

\section{Conclusions} \label{sec:conclusion} 
We have investigated the manipulation of light scattered by a single nanoparticle with optical wavefront shaping (WFS), {\color{black}as a complement to the traditional WFS of ensembles of many particles}. 
Our results show that wavefront shaping can serve as an efficient knob to turn on different highly scattering eigenchannels intrinsic to the scatterer. 
These highly scattering channels are found to be related to the different resonant leaky modes of the scatterer, indicating that manipulation of nanoparticle scattering through WFS offers a possible route toward controlling light-matter interaction with individual nanoparticles.
Moreover, we have found that both short range and long range spectral correlations exist for the highly scattering eigenchannels of the scatterer, and the correlations are related to different types of resonant leaky modes of the nanoparticle. 
Our results demonstrate that optical WFS is not only a powerful method for media composed of large numbers of nanoparticles, but also an efficient method to manipulate scattering by a \textit{single} nanoparticle. 
{\color{black}As a result, we propose novel opportunities with light-matter interactions and with ultrafast phenomena that could profit from our study. }
Thus, our results offer new perspectives in the understanding of complex scattering processes involving nanoparticles and other resonant optical systems.

\section*{Appendix} \label{sec:Appendix A} 

\begin{figure}[htb]
	\centering
	\includegraphics[width=0.48\textwidth]{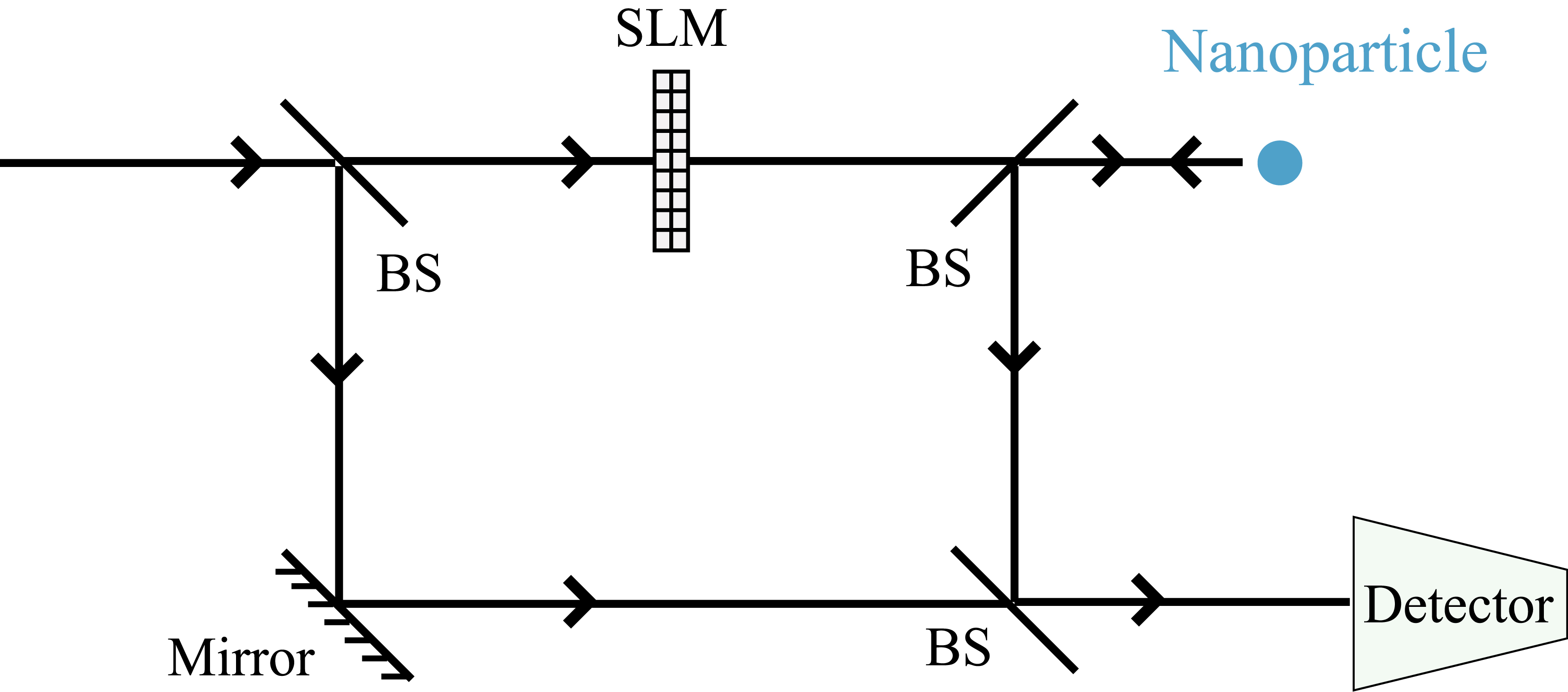}
	\caption{
 {\color{black}Schematic diagram of an optical setup to measure the scattering matrix of a single nanoparticle in backscatter directions and for conducting WFS to selectively coupling light to a backscattered eigenchannel. }
 \label{fig:Fig4_Diagram}}
\end{figure}

{\color{black}Figure~\ref{fig:Fig4_Diagram} shows a schematic diagram of a possible setup to measure the scattering matrix of a single nanoparticle in backscattering (as discussed here) and to couple light to an eigenchannel in a backscattering direction by optical wavefront shaping. 
The incident light is sent to an interferometer with two arms.
In the probe arm, the light beam is sent to a nanoparticle after being spatially modulated with an SLM~\cite{vellekoop2007focusing, popoff2010measuring, mosk2012controlling}, and the backscattered light is directed to the detector located at the far field. 
In the reference arm, the light beam propagates freely to the detector.
For simplicity, we ignored the objective lens before the nanoparticle and the relay lens in the setup. 
By measuring the phase-shifting holograms controlled by the SLM~\cite{popoff2010measuring}, the backscattering matrix of the nanoparticle can be retrieved.
To selectively coupling the eigenchannels of the nanoparticle, one can block the reference arm, and control the wavefront of incident light by the SLM. 
} 

\vspace{9pt}
\section*{Acknowledgments}
We thank Ad Lagendijk, Allard Mosk, and Paul Urbach for encouragements and insightful discussions. 
PH acknowledges support from the Fundamental Research Funds for the Central Universities (ZYGX2020J010), and the Open Project Funding of the MOE Key Laboratory of Weak-Light Nonlinear Photonics (OS22-1).
WLV acknowledges support by the NWO-TTW program P15-36 ”Free-form scattering optics” (FFSO, in collaboration with TUE, TUD, and ASML, Demcon, Lumileds, Schott, Signify, TNO), NWO program "Photonic band gaps in quasi-crystals" (PI: Alfons van Blaaderen), and the MESA+ Institute section Applied Nanophotonics (ANP).

\vspace{9pt}
\bibliography{literature}

\begin{thebibliography}{79}%
\makeatletter
\providecommand \@ifxundefined [1]{%
 \@ifx{#1\undefined}
}%
\providecommand \@ifnum [1]{%
 \ifnum #1\expandafter \@firstoftwo
 \else \expandafter \@secondoftwo
 \fi
}%
\providecommand \@ifx [1]{%
 \ifx #1\expandafter \@firstoftwo
 \else \expandafter \@secondoftwo
 \fi
}%
\providecommand \natexlab [1]{#1}%
\providecommand \enquote  [1]{``#1''}%
\providecommand \bibnamefont  [1]{#1}%
\providecommand \bibfnamefont [1]{#1}%
\providecommand \citenamefont [1]{#1}%
\providecommand \href@noop [0]{\@secondoftwo}%
\providecommand \href [0]{\begingroup \@sanitize@url \@href}%
\providecommand \@href[1]{\@@startlink{#1}\@@href}%
\providecommand \@@href[1]{\endgroup#1\@@endlink}%
\providecommand \@sanitize@url [0]{\catcode `\\12\catcode `\$12\catcode
  `\&12\catcode `\#12\catcode `\^12\catcode `\_12\catcode `\%12\relax}%
\providecommand \@@startlink[1]{}%
\providecommand \@@endlink[0]{}%
\providecommand \url  [0]{\begingroup\@sanitize@url \@url }%
\providecommand \@url [1]{\endgroup\@href {#1}{\urlprefix }}%
\providecommand \urlprefix  [0]{URL }%
\providecommand \Eprint [0]{\href }%
\providecommand \doibase [0]{https://doi.org/}%
\providecommand \selectlanguage [0]{\@gobble}%
\providecommand \bibinfo  [0]{\@secondoftwo}%
\providecommand \bibfield  [0]{\@secondoftwo}%
\providecommand \translation [1]{[#1]}%
\providecommand \BibitemOpen [0]{}%
\providecommand \bibitemStop [0]{}%
\providecommand \bibitemNoStop [0]{.\EOS\space}%
\providecommand \EOS [0]{\spacefactor3000\relax}%
\providecommand \BibitemShut  [1]{\csname bibitem#1\endcsname}%
\let\auto@bib@innerbib\@empty
\bibitem [{\citenamefont {Akkermans}\ and\ \citenamefont
  {Montambaux}(2007)}]{Akkermans2007BOOK}%
  \BibitemOpen
  \bibfield  {author} {\bibinfo {author} {\bibfnamefont {E.}~\bibnamefont
  {Akkermans}}\ and\ \bibinfo {author} {\bibfnamefont {G.}~\bibnamefont
  {Montambaux}},\ }\href@noop {} {\emph {\bibinfo {title} {Mesoscopic physics
  of electrons and photons}}}\ (\bibinfo  {publisher} {Cambridge university
  press},\ \bibinfo {year} {2007})\BibitemShut {NoStop}%
\bibitem [{\citenamefont {Joannopoulos}\ \emph {et~al.}(2008)\citenamefont
  {Joannopoulos}, \citenamefont {Johnson}, \citenamefont {Winn},\ and\
  \citenamefont {Meade}}]{joannopoulos2008molding}%
  \BibitemOpen
  \bibfield  {author} {\bibinfo {author} {\bibfnamefont {J.~D.}\ \bibnamefont
  {Joannopoulos}}, \bibinfo {author} {\bibfnamefont {S.~G.}\ \bibnamefont
  {Johnson}}, \bibinfo {author} {\bibfnamefont {J.~N.}\ \bibnamefont {Winn}},\
  and\ \bibinfo {author} {\bibfnamefont {R.~D.}\ \bibnamefont {Meade}},\
  }\href@noop {} {\emph {\bibinfo {title} {Photonic Crystals: molding the flow
  of light}}}\ (\bibinfo  {publisher} {Princeton Univ. Press, Princeton},\
  \bibinfo {year} {2008})\BibitemShut {NoStop}%
\bibitem [{\citenamefont {Lourtioz}\ \emph {et~al.}(2008)\citenamefont
  {Lourtioz}, \citenamefont {Benisty}, \citenamefont {Berger}, \citenamefont
  {G{\'e}rard}, \citenamefont {Maystre},\ and\ \citenamefont
  {Tchelnokov}}]{Lourtioz2008Book}%
  \BibitemOpen
  \bibinfo {editor} {\bibfnamefont {J.-M.}\ \bibnamefont {Lourtioz}}, \bibinfo
  {editor} {\bibfnamefont {H.}~\bibnamefont {Benisty}}, \bibinfo {editor}
  {\bibfnamefont {V.}~\bibnamefont {Berger}}, \bibinfo {editor} {\bibfnamefont
  {J.-M.}\ \bibnamefont {G{\'e}rard}}, \bibinfo {editor} {\bibfnamefont
  {D.}~\bibnamefont {Maystre}},\ and\ \bibinfo {editor} {\bibfnamefont
  {A.}~\bibnamefont {Tchelnokov}},\ eds.,\ \href@noop {} {\emph {\bibinfo
  {title} {Photonic Crystals: Towards Nanoscale Photonic Devices}}}\ (\bibinfo
  {publisher} {Springer, New York NY},\ \bibinfo {year} {2008})\BibitemShut
  {NoStop}%
\bibitem [{\citenamefont {Ghulinyan}\ and\ \citenamefont
  {Pavesi}(2015)}]{Ghulinyan2015Book}%
  \BibitemOpen
  \bibinfo {editor} {\bibfnamefont {M.}~\bibnamefont {Ghulinyan}}\ and\
  \bibinfo {editor} {\bibfnamefont {L.}~\bibnamefont {Pavesi}},\ eds.,\
  \href@noop {} {\emph {\bibinfo {title} {Light Localisation and Lasing: Random
  and Pseudorandom Photonic Structures}}}\ (\bibinfo  {publisher} {Cambridge
  Univ. Press, Cambridge UK},\ \bibinfo {year} {2015})\BibitemShut {NoStop}%
\bibitem [{\citenamefont {Barnes}\ \emph {et~al.}(2003)\citenamefont {Barnes},
  \citenamefont {Dereux},\ and\ \citenamefont {Ebbesen}}]{Barnes2003Nature}%
  \BibitemOpen
  \bibfield  {author} {\bibinfo {author} {\bibfnamefont {W.~L.}\ \bibnamefont
  {Barnes}}, \bibinfo {author} {\bibfnamefont {A.}~\bibnamefont {Dereux}},\
  and\ \bibinfo {author} {\bibfnamefont {T.~W.}\ \bibnamefont {Ebbesen}},\
  }\bibfield  {title} {\bibinfo {title} {Surface plasmon subwavelength
  optics},\ }\href@noop {} {\bibfield  {journal} {\bibinfo  {journal} {Nature}\
  }\textbf {\bibinfo {volume} {424}},\ \bibinfo {pages} {824} (\bibinfo {year}
  {2003})}\BibitemShut {NoStop}%
\bibitem [{\citenamefont {Soukoulis}\ and\ \citenamefont
  {Wegener}(2011)}]{Soukoulis2011NatPhoton}%
  \BibitemOpen
  \bibfield  {author} {\bibinfo {author} {\bibfnamefont {C.~M.}\ \bibnamefont
  {Soukoulis}}\ and\ \bibinfo {author} {\bibfnamefont {M.}~\bibnamefont
  {Wegener}},\ }\bibfield  {title} {\bibinfo {title} {Past achievements and
  future challenges in the development of three-dimensional photonic
  metamaterials},\ }\href@noop {} {\bibfield  {journal} {\bibinfo  {journal}
  {Nat. Photon.}\ }\textbf {\bibinfo {volume} {5}},\ \bibinfo {pages} {523}
  (\bibinfo {year} {2011})}\BibitemShut {NoStop}%
\bibitem [{\citenamefont {Yu}\ and\ \citenamefont
  {Capasso}(2014)}]{yu2014flat}%
  \BibitemOpen
  \bibfield  {author} {\bibinfo {author} {\bibfnamefont {N.}~\bibnamefont
  {Yu}}\ and\ \bibinfo {author} {\bibfnamefont {F.}~\bibnamefont {Capasso}},\
  }\bibfield  {title} {\bibinfo {title} {Flat optics with designer
  metasurfaces},\ }\href@noop {} {\bibfield  {journal} {\bibinfo  {journal}
  {Nat. Mater.}\ }\textbf {\bibinfo {volume} {13}},\ \bibinfo {pages} {139}
  (\bibinfo {year} {2014})}\BibitemShut {NoStop}%
\bibitem [{\citenamefont {Lagendijk}\ \emph {et~al.}(2009)\citenamefont
  {Lagendijk}, \citenamefont {Van~Tiggelen},\ and\ \citenamefont
  {Wiersma}}]{Lagendijk2009PhysToday}%
  \BibitemOpen
  \bibfield  {author} {\bibinfo {author} {\bibfnamefont {A.}~\bibnamefont
  {Lagendijk}}, \bibinfo {author} {\bibfnamefont {B.}~\bibnamefont
  {Van~Tiggelen}},\ and\ \bibinfo {author} {\bibfnamefont {D.~S.}\ \bibnamefont
  {Wiersma}},\ }\bibfield  {title} {\bibinfo {title} {Fifty years of anderson
  localization},\ }\href@noop {} {\bibfield  {journal} {\bibinfo  {journal}
  {Phys. Today}\ }\textbf {\bibinfo {volume} {62}},\ \bibinfo {pages} {24}
  (\bibinfo {year} {2009})}\BibitemShut {NoStop}%
\bibitem [{\citenamefont {Wiersma}(2013)}]{Wiersma2013NatPhot}%
  \BibitemOpen
  \bibfield  {author} {\bibinfo {author} {\bibfnamefont {D.~S.}\ \bibnamefont
  {Wiersma}},\ }\bibfield  {title} {\bibinfo {title} {Disordered photonics},\
  }\href@noop {} {\bibfield  {journal} {\bibinfo  {journal} {Nat. Photon.}\
  }\textbf {\bibinfo {volume} {7}},\ \bibinfo {pages} {188} (\bibinfo {year}
  {2013})}\BibitemShut {NoStop}%
\bibitem [{\citenamefont {Segev}\ \emph {et~al.}(2013)\citenamefont {Segev},
  \citenamefont {Silberberg},\ and\ \citenamefont
  {Christodoulides}}]{Segev2013NatPhot}%
  \BibitemOpen
  \bibfield  {author} {\bibinfo {author} {\bibfnamefont {M.}~\bibnamefont
  {Segev}}, \bibinfo {author} {\bibfnamefont {Y.}~\bibnamefont {Silberberg}},\
  and\ \bibinfo {author} {\bibfnamefont {D.~N.}\ \bibnamefont
  {Christodoulides}},\ }\bibfield  {title} {\bibinfo {title} {Anderson
  localization of light},\ }\href@noop {} {\bibfield  {journal} {\bibinfo
  {journal} {Nat. Photon.}\ }\textbf {\bibinfo {volume} {7}},\ \bibinfo {pages}
  {197} (\bibinfo {year} {2013})}\BibitemShut {NoStop}%
\bibitem [{\citenamefont {Skipetrov}\ and\ \citenamefont
  {Page}(2016)}]{Skipetrov2016NJP}%
  \BibitemOpen
  \bibfield  {author} {\bibinfo {author} {\bibfnamefont {S.}~\bibnamefont
  {Skipetrov}}\ and\ \bibinfo {author} {\bibfnamefont {J.~H.}\ \bibnamefont
  {Page}},\ }\bibfield  {title} {\bibinfo {title} {Red light for anderson
  localization},\ }\href@noop {} {\bibfield  {journal} {\bibinfo  {journal}
  {New J. Phys.}\ }\textbf {\bibinfo {volume} {18}},\ \bibinfo {pages} {021001}
  (\bibinfo {year} {2016})}\BibitemShut {NoStop}%
\bibitem [{\citenamefont {Skipetrov}(2020)}]{Skipetrov2020PRB}%
  \BibitemOpen
  \bibfield  {author} {\bibinfo {author} {\bibfnamefont {S.}~\bibnamefont
  {Skipetrov}},\ }\bibfield  {title} {\bibinfo {title} {Localization of light
  in a three-dimensional disordered crystal of atoms},\ }\href@noop {}
  {\bibfield  {journal} {\bibinfo  {journal} {Phys. Rev. B}\ }\textbf {\bibinfo
  {volume} {102}},\ \bibinfo {pages} {134206} (\bibinfo {year}
  {2020})}\BibitemShut {NoStop}%
\bibitem [{\citenamefont {{van Albada, Meint P and Lagendijk,
  Ad}}(1985)}]{Vanalbada1985PRL}%
  \BibitemOpen
  \bibfield  {author} {\bibinfo {author} {\bibnamefont {{van Albada, Meint P
  and Lagendijk, Ad}}},\ }\bibfield  {title} {\bibinfo {title} {Observation of
  weak localization of light in a random medium},\ }\href@noop {} {\bibfield
  {journal} {\bibinfo  {journal} {Phys. Rev. Lett.}\ }\textbf {\bibinfo
  {volume} {55}},\ \bibinfo {pages} {2692} (\bibinfo {year}
  {1985})}\BibitemShut {NoStop}%
\bibitem [{\citenamefont {Wolf}\ and\ \citenamefont
  {Maret}(1985)}]{Wolf1985PRL}%
  \BibitemOpen
  \bibfield  {author} {\bibinfo {author} {\bibfnamefont {P.-E.}\ \bibnamefont
  {Wolf}}\ and\ \bibinfo {author} {\bibfnamefont {G.}~\bibnamefont {Maret}},\
  }\bibfield  {title} {\bibinfo {title} {Weak localization and coherent
  backscattering of photons in disordered media},\ }\href@noop {} {\bibfield
  {journal} {\bibinfo  {journal} {Phys. Rev. Lett.}\ }\textbf {\bibinfo
  {volume} {55}},\ \bibinfo {pages} {2696} (\bibinfo {year}
  {1985})}\BibitemShut {NoStop}%
\bibitem [{\citenamefont {Fazio}\ \emph {et~al.}(2017)\citenamefont {Fazio},
  \citenamefont {Irrera}, \citenamefont {Pirotta}, \citenamefont {D'Andrea},
  \citenamefont {Del~Sorbo}, \citenamefont {Faro}, \citenamefont {Gucciardi},
  \citenamefont {Iati}, \citenamefont {Saija}, \citenamefont {Patrini},
  \citenamefont {Musumeci}, \citenamefont {Vasi}, \citenamefont {Wiersma},
  \citenamefont {Galli},\ and\ \citenamefont {Priolo}}]{Fazio2017NatPhot}%
  \BibitemOpen
  \bibfield  {author} {\bibinfo {author} {\bibfnamefont {B.}~\bibnamefont
  {Fazio}}, \bibinfo {author} {\bibfnamefont {A.}~\bibnamefont {Irrera}},
  \bibinfo {author} {\bibfnamefont {S.}~\bibnamefont {Pirotta}}, \bibinfo
  {author} {\bibfnamefont {C.}~\bibnamefont {D'Andrea}}, \bibinfo {author}
  {\bibfnamefont {S.}~\bibnamefont {Del~Sorbo}}, \bibinfo {author}
  {\bibfnamefont {M.~J.~L.}\ \bibnamefont {Faro}}, \bibinfo {author}
  {\bibfnamefont {P.~G.}\ \bibnamefont {Gucciardi}}, \bibinfo {author}
  {\bibfnamefont {M.~A.}\ \bibnamefont {Iati}}, \bibinfo {author}
  {\bibfnamefont {R.}~\bibnamefont {Saija}}, \bibinfo {author} {\bibfnamefont
  {M.}~\bibnamefont {Patrini}}, \bibinfo {author} {\bibfnamefont
  {P.}~\bibnamefont {Musumeci}}, \bibinfo {author} {\bibfnamefont {C.~S.}\
  \bibnamefont {Vasi}}, \bibinfo {author} {\bibfnamefont {D.~S.}\ \bibnamefont
  {Wiersma}}, \bibinfo {author} {\bibfnamefont {M.}~\bibnamefont {Galli}},\
  and\ \bibinfo {author} {\bibfnamefont {F.}~\bibnamefont {Priolo}},\
  }\bibfield  {title} {\bibinfo {title} {{Coherent backscattering of Raman
  light}},\ }\href {https://doi.org/https://doi.org/10.1038/nphoton.2016.278}
  {\bibfield  {journal} {\bibinfo  {journal} {Nat. Photon.}\ }\textbf {\bibinfo
  {volume} {11}},\ \bibinfo {pages} {170} (\bibinfo {year} {2017})}\BibitemShut
  {NoStop}%
\bibitem [{\citenamefont {Cao}\ \emph {et~al.}(1999)\citenamefont {Cao},
  \citenamefont {Zhao}, \citenamefont {Ho}, \citenamefont {Seelig},
  \citenamefont {Wang},\ and\ \citenamefont {Chang}}]{cao1999random}%
  \BibitemOpen
  \bibfield  {author} {\bibinfo {author} {\bibfnamefont {H.}~\bibnamefont
  {Cao}}, \bibinfo {author} {\bibfnamefont {Y.}~\bibnamefont {Zhao}}, \bibinfo
  {author} {\bibfnamefont {S.}~\bibnamefont {Ho}}, \bibinfo {author}
  {\bibfnamefont {E.}~\bibnamefont {Seelig}}, \bibinfo {author} {\bibfnamefont
  {Q.}~\bibnamefont {Wang}},\ and\ \bibinfo {author} {\bibfnamefont
  {R.}~\bibnamefont {Chang}},\ }\bibfield  {title} {\bibinfo {title} {Random
  laser action in semiconductor powder},\ }\href@noop {} {\bibfield  {journal}
  {\bibinfo  {journal} {Phys. Rev. Lett.}\ }\textbf {\bibinfo {volume} {82}},\
  \bibinfo {pages} {2278} (\bibinfo {year} {1999})}\BibitemShut {NoStop}%
\bibitem [{\citenamefont {Fallert}\ \emph {et~al.}(2009)\citenamefont
  {Fallert}, \citenamefont {Dietz}, \citenamefont {Sartor}, \citenamefont
  {Schneider}, \citenamefont {Klingshirn},\ and\ \citenamefont
  {Kalt}}]{fallert2009co}%
  \BibitemOpen
  \bibfield  {author} {\bibinfo {author} {\bibfnamefont {J.}~\bibnamefont
  {Fallert}}, \bibinfo {author} {\bibfnamefont {R.~J.}\ \bibnamefont {Dietz}},
  \bibinfo {author} {\bibfnamefont {J.}~\bibnamefont {Sartor}}, \bibinfo
  {author} {\bibfnamefont {D.}~\bibnamefont {Schneider}}, \bibinfo {author}
  {\bibfnamefont {C.}~\bibnamefont {Klingshirn}},\ and\ \bibinfo {author}
  {\bibfnamefont {H.}~\bibnamefont {Kalt}},\ }\bibfield  {title} {\bibinfo
  {title} {Co-existence of strongly and weakly localized random laser modes},\
  }\href@noop {} {\bibfield  {journal} {\bibinfo  {journal} {Nat. Photon.}\
  }\textbf {\bibinfo {volume} {3}},\ \bibinfo {pages} {279} (\bibinfo {year}
  {2009})}\BibitemShut {NoStop}%
\bibitem [{\citenamefont {Bertolotti}\ \emph {et~al.}(2012)\citenamefont
  {Bertolotti}, \citenamefont {Van~Putten}, \citenamefont {Blum}, \citenamefont
  {Lagendijk}, \citenamefont {Vos},\ and\ \citenamefont
  {Mosk}}]{bertolotti2012non}%
  \BibitemOpen
  \bibfield  {author} {\bibinfo {author} {\bibfnamefont {J.}~\bibnamefont
  {Bertolotti}}, \bibinfo {author} {\bibfnamefont {E.~G.}\ \bibnamefont
  {Van~Putten}}, \bibinfo {author} {\bibfnamefont {C.}~\bibnamefont {Blum}},
  \bibinfo {author} {\bibfnamefont {A.}~\bibnamefont {Lagendijk}}, \bibinfo
  {author} {\bibfnamefont {W.~L.}\ \bibnamefont {Vos}},\ and\ \bibinfo {author}
  {\bibfnamefont {A.~P.}\ \bibnamefont {Mosk}},\ }\bibfield  {title} {\bibinfo
  {title} {Non-invasive imaging through opaque scattering layers},\ }\href@noop
  {} {\bibfield  {journal} {\bibinfo  {journal} {Nature}\ }\textbf {\bibinfo
  {volume} {491}},\ \bibinfo {pages} {232} (\bibinfo {year}
  {2012})}\BibitemShut {NoStop}%
\bibitem [{\citenamefont {Katz}\ \emph {et~al.}(2014)\citenamefont {Katz},
  \citenamefont {Heidmann}, \citenamefont {Fink},\ and\ \citenamefont
  {Gigan}}]{katz2014non}%
  \BibitemOpen
  \bibfield  {author} {\bibinfo {author} {\bibfnamefont {O.}~\bibnamefont
  {Katz}}, \bibinfo {author} {\bibfnamefont {P.}~\bibnamefont {Heidmann}},
  \bibinfo {author} {\bibfnamefont {M.}~\bibnamefont {Fink}},\ and\ \bibinfo
  {author} {\bibfnamefont {S.}~\bibnamefont {Gigan}},\ }\bibfield  {title}
  {\bibinfo {title} {Non-invasive single-shot imaging through scattering layers
  and around corners via speckle correlations},\ }\href@noop {} {\bibfield
  {journal} {\bibinfo  {journal} {Nat. Photon.}\ }\textbf {\bibinfo {volume}
  {8}},\ \bibinfo {pages} {784} (\bibinfo {year} {2014})}\BibitemShut {NoStop}%
\bibitem [{\citenamefont {Hong}(2018)}]{Hong2018APL}%
  \BibitemOpen
  \bibfield  {author} {\bibinfo {author} {\bibfnamefont {P.}~\bibnamefont
  {Hong}},\ }\bibfield  {title} {\bibinfo {title} {Two-photon imaging assisted
  by a thin dynamic scattering layer},\ }\href@noop {} {\bibfield  {journal}
  {\bibinfo  {journal} {Appl. Phys. Lett.}\ }\textbf {\bibinfo {volume}
  {113}},\ \bibinfo {pages} {101109} (\bibinfo {year} {2018})}\BibitemShut
  {NoStop}%
\bibitem [{\citenamefont {Sarma}\ \emph {et~al.}(2016)\citenamefont {Sarma},
  \citenamefont {Yamilov}, \citenamefont {Petrenko}, \citenamefont {Bromberg},\
  and\ \citenamefont {Cao}}]{sarma2016control}%
  \BibitemOpen
  \bibfield  {author} {\bibinfo {author} {\bibfnamefont {R.}~\bibnamefont
  {Sarma}}, \bibinfo {author} {\bibfnamefont {A.~G.}\ \bibnamefont {Yamilov}},
  \bibinfo {author} {\bibfnamefont {S.}~\bibnamefont {Petrenko}}, \bibinfo
  {author} {\bibfnamefont {Y.}~\bibnamefont {Bromberg}},\ and\ \bibinfo
  {author} {\bibfnamefont {H.}~\bibnamefont {Cao}},\ }\bibfield  {title}
  {\bibinfo {title} {Control of energy density inside a disordered medium by
  coupling to open or closed channels},\ }\href@noop {} {\bibfield  {journal}
  {\bibinfo  {journal} {Phys. Rev. Lett.}\ }\textbf {\bibinfo {volume} {117}},\
  \bibinfo {pages} {086803} (\bibinfo {year} {2016})}\BibitemShut {NoStop}%
\bibitem [{\citenamefont {Durduran}\ \emph {et~al.}(2010)\citenamefont
  {Durduran}, \citenamefont {Choe}, \citenamefont {Baker},\ and\ \citenamefont
  {Yodh}}]{Durduran2010RepProgPhys}%
  \BibitemOpen
  \bibfield  {author} {\bibinfo {author} {\bibfnamefont {T.}~\bibnamefont
  {Durduran}}, \bibinfo {author} {\bibfnamefont {R.}~\bibnamefont {Choe}},
  \bibinfo {author} {\bibfnamefont {W.~B.}\ \bibnamefont {Baker}},\ and\
  \bibinfo {author} {\bibfnamefont {A.~G.}\ \bibnamefont {Yodh}},\ }\bibfield
  {title} {\bibinfo {title} {Diffuse optics for tissue monitoring and
  tomography},\ }\href@noop {} {\bibfield  {journal} {\bibinfo  {journal} {Rep.
  Prog. Phys}\ }\textbf {\bibinfo {volume} {73}},\ \bibinfo {pages} {076701}
  (\bibinfo {year} {2010})}\BibitemShut {NoStop}%
\bibitem [{\citenamefont {Mariani}\ \emph {et~al.}(2018)\citenamefont
  {Mariani}, \citenamefont {L\"{o}ffler}, \citenamefont {Aas}, \citenamefont
  {Ojambati}, \citenamefont {Hong}, \citenamefont {Vos},\ and\ \citenamefont
  {van Exter}}]{Mariani2018OE}%
  \BibitemOpen
  \bibfield  {author} {\bibinfo {author} {\bibfnamefont {F.}~\bibnamefont
  {Mariani}}, \bibinfo {author} {\bibfnamefont {W.}~\bibnamefont
  {L\"{o}ffler}}, \bibinfo {author} {\bibfnamefont {M.}~\bibnamefont {Aas}},
  \bibinfo {author} {\bibfnamefont {O.~S.}\ \bibnamefont {Ojambati}}, \bibinfo
  {author} {\bibfnamefont {P.}~\bibnamefont {Hong}}, \bibinfo {author}
  {\bibfnamefont {W.~L.}\ \bibnamefont {Vos}},\ and\ \bibinfo {author}
  {\bibfnamefont {M.~P.}\ \bibnamefont {van Exter}},\ }\bibfield  {title}
  {\bibinfo {title} {Scattering media characterization with phase-only
  wavefront modulation},\ }\href {https://doi.org/10.1364/OE.26.002369}
  {\bibfield  {journal} {\bibinfo  {journal} {Opt. Express}\ }\textbf {\bibinfo
  {volume} {26}},\ \bibinfo {pages} {2369} (\bibinfo {year}
  {2018})}\BibitemShut {NoStop}%
\bibitem [{\citenamefont {Vellekoop}\ and\ \citenamefont
  {Mosk}(2007)}]{vellekoop2007focusing}%
  \BibitemOpen
  \bibfield  {author} {\bibinfo {author} {\bibfnamefont {I.~M.}\ \bibnamefont
  {Vellekoop}}\ and\ \bibinfo {author} {\bibfnamefont {A.}~\bibnamefont
  {Mosk}},\ }\bibfield  {title} {\bibinfo {title} {Focusing coherent light
  through opaque strongly scattering media},\ }\href@noop {} {\bibfield
  {journal} {\bibinfo  {journal} {Opt. Lett.}\ }\textbf {\bibinfo {volume}
  {32}},\ \bibinfo {pages} {2309} (\bibinfo {year} {2007})}\BibitemShut
  {NoStop}%
\bibitem [{\citenamefont {Popoff}\ \emph {et~al.}(2010)\citenamefont {Popoff},
  \citenamefont {Lerosey}, \citenamefont {Carminati}, \citenamefont {Fink},
  \citenamefont {Boccara},\ and\ \citenamefont {Gigan}}]{popoff2010measuring}%
  \BibitemOpen
  \bibfield  {author} {\bibinfo {author} {\bibfnamefont {S.~M.}\ \bibnamefont
  {Popoff}}, \bibinfo {author} {\bibfnamefont {G.}~\bibnamefont {Lerosey}},
  \bibinfo {author} {\bibfnamefont {R.}~\bibnamefont {Carminati}}, \bibinfo
  {author} {\bibfnamefont {M.}~\bibnamefont {Fink}}, \bibinfo {author}
  {\bibfnamefont {A.~C.}\ \bibnamefont {Boccara}},\ and\ \bibinfo {author}
  {\bibfnamefont {S.}~\bibnamefont {Gigan}},\ }\bibfield  {title} {\bibinfo
  {title} {Measuring the transmission matrix in optics: an approach to the
  study and control of light propagation in disordered media},\ }\href@noop {}
  {\bibfield  {journal} {\bibinfo  {journal} {Phys. Rev. Lett.}\ }\textbf
  {\bibinfo {volume} {104}},\ \bibinfo {pages} {100601} (\bibinfo {year}
  {2010})}\BibitemShut {NoStop}%
\bibitem [{\citenamefont {Mosk}\ \emph {et~al.}(2012)\citenamefont {Mosk},
  \citenamefont {Lagendijk}, \citenamefont {Lerosey},\ and\ \citenamefont
  {Fink}}]{mosk2012controlling}%
  \BibitemOpen
  \bibfield  {author} {\bibinfo {author} {\bibfnamefont {A.~P.}\ \bibnamefont
  {Mosk}}, \bibinfo {author} {\bibfnamefont {A.}~\bibnamefont {Lagendijk}},
  \bibinfo {author} {\bibfnamefont {G.}~\bibnamefont {Lerosey}},\ and\ \bibinfo
  {author} {\bibfnamefont {M.}~\bibnamefont {Fink}},\ }\bibfield  {title}
  {\bibinfo {title} {Controlling waves in space and time for imaging and
  focusing in complex media},\ }\href@noop {} {\bibfield  {journal} {\bibinfo
  {journal} {Nat. Photon.}\ }\textbf {\bibinfo {volume} {6}},\ \bibinfo {pages}
  {283} (\bibinfo {year} {2012})}\BibitemShut {NoStop}%
\bibitem [{\citenamefont {Vellekoop}\ \emph {et~al.}(2010)\citenamefont
  {Vellekoop}, \citenamefont {Lagendijk},\ and\ \citenamefont
  {Mosk}}]{vellekoop2010exploiting}%
  \BibitemOpen
  \bibfield  {author} {\bibinfo {author} {\bibfnamefont {I.~M.}\ \bibnamefont
  {Vellekoop}}, \bibinfo {author} {\bibfnamefont {A.}~\bibnamefont
  {Lagendijk}},\ and\ \bibinfo {author} {\bibfnamefont {A.}~\bibnamefont
  {Mosk}},\ }\bibfield  {title} {\bibinfo {title} {Exploiting disorder for
  perfect focusing},\ }\href@noop {} {\bibfield  {journal} {\bibinfo  {journal}
  {Nat. Photon.}\ }\textbf {\bibinfo {volume} {4}},\ \bibinfo {pages} {320}
  (\bibinfo {year} {2010})}\BibitemShut {NoStop}%
\bibitem [{\citenamefont {van Putten}\ \emph {et~al.}(2011)\citenamefont {van
  Putten}, \citenamefont {Akbulut}, \citenamefont {Bertolotti}, \citenamefont
  {Vos}, \citenamefont {Lagendijk},\ and\ \citenamefont
  {Mosk}}]{vanputten2011scattering}%
  \BibitemOpen
  \bibfield  {author} {\bibinfo {author} {\bibfnamefont {E.~G.}\ \bibnamefont
  {van Putten}}, \bibinfo {author} {\bibfnamefont {D.}~\bibnamefont {Akbulut}},
  \bibinfo {author} {\bibfnamefont {J.}~\bibnamefont {Bertolotti}}, \bibinfo
  {author} {\bibfnamefont {W.~L.}\ \bibnamefont {Vos}}, \bibinfo {author}
  {\bibfnamefont {A.}~\bibnamefont {Lagendijk}},\ and\ \bibinfo {author}
  {\bibfnamefont {A.}~\bibnamefont {Mosk}},\ }\bibfield  {title} {\bibinfo
  {title} {Scattering lens resolves sub-100 nm structures with visible light},\
  }\href@noop {} {\bibfield  {journal} {\bibinfo  {journal} {Phys. Rev. Lett.}\
  }\textbf {\bibinfo {volume} {106}},\ \bibinfo {pages} {193905} (\bibinfo
  {year} {2011})}\BibitemShut {NoStop}%
\bibitem [{\citenamefont {de~Aguiar}\ \emph {et~al.}(2017)\citenamefont
  {de~Aguiar}, \citenamefont {Gigan},\ and\ \citenamefont
  {Brasselet}}]{deaguiar2017polarization}%
  \BibitemOpen
  \bibfield  {author} {\bibinfo {author} {\bibfnamefont {H.~B.}\ \bibnamefont
  {de~Aguiar}}, \bibinfo {author} {\bibfnamefont {S.}~\bibnamefont {Gigan}},\
  and\ \bibinfo {author} {\bibfnamefont {S.}~\bibnamefont {Brasselet}},\
  }\bibfield  {title} {\bibinfo {title} {Polarization recovery through
  scattering media},\ }\href@noop {} {\bibfield  {journal} {\bibinfo  {journal}
  {Sci. Adv.}\ }\textbf {\bibinfo {volume} {3}},\ \bibinfo {pages} {e1600743}
  (\bibinfo {year} {2017})}\BibitemShut {NoStop}%
\bibitem [{\citenamefont {Yeminy}\ and\ \citenamefont
  {Katz}(2021)}]{yeminy2021guidestar}%
  \BibitemOpen
  \bibfield  {author} {\bibinfo {author} {\bibfnamefont {T.}~\bibnamefont
  {Yeminy}}\ and\ \bibinfo {author} {\bibfnamefont {O.}~\bibnamefont {Katz}},\
  }\bibfield  {title} {\bibinfo {title} {Guidestar-free image-guided wavefront
  shaping},\ }\href@noop {} {\bibfield  {journal} {\bibinfo  {journal} {Sci.
  Adv.}\ }\textbf {\bibinfo {volume} {7}},\ \bibinfo {pages} {eabf5364}
  (\bibinfo {year} {2021})}\BibitemShut {NoStop}%
\bibitem [{\citenamefont {Pai}\ \emph {et~al.}(2021)\citenamefont {Pai},
  \citenamefont {Bosch}, \citenamefont {K{\"u}hmayer}, \citenamefont {Rotter},\
  and\ \citenamefont {Mosk}}]{pai2021scattering}%
  \BibitemOpen
  \bibfield  {author} {\bibinfo {author} {\bibfnamefont {P.}~\bibnamefont
  {Pai}}, \bibinfo {author} {\bibfnamefont {J.}~\bibnamefont {Bosch}}, \bibinfo
  {author} {\bibfnamefont {M.}~\bibnamefont {K{\"u}hmayer}}, \bibinfo {author}
  {\bibfnamefont {S.}~\bibnamefont {Rotter}},\ and\ \bibinfo {author}
  {\bibfnamefont {A.~P.}\ \bibnamefont {Mosk}},\ }\bibfield  {title} {\bibinfo
  {title} {Scattering invariant modes of light in complex media},\ }\href@noop
  {} {\bibfield  {journal} {\bibinfo  {journal} {Nat. Photon.}\ }\textbf
  {\bibinfo {volume} {15}},\ \bibinfo {pages} {431} (\bibinfo {year}
  {2021})}\BibitemShut {NoStop}%
\bibitem [{\citenamefont {Choi}\ \emph {et~al.}(2011)\citenamefont {Choi},
  \citenamefont {Mosk}, \citenamefont {Park},\ and\ \citenamefont
  {Choi}}]{choi2011transmission}%
  \BibitemOpen
  \bibfield  {author} {\bibinfo {author} {\bibfnamefont {W.}~\bibnamefont
  {Choi}}, \bibinfo {author} {\bibfnamefont {A.~P.}\ \bibnamefont {Mosk}},
  \bibinfo {author} {\bibfnamefont {Q.-H.}\ \bibnamefont {Park}},\ and\
  \bibinfo {author} {\bibfnamefont {W.}~\bibnamefont {Choi}},\ }\bibfield
  {title} {\bibinfo {title} {Transmission eigenchannels in a disordered
  medium},\ }\href@noop {} {\bibfield  {journal} {\bibinfo  {journal} {Phys.
  Rev. B}\ }\textbf {\bibinfo {volume} {83}},\ \bibinfo {pages} {134207}
  (\bibinfo {year} {2011})}\BibitemShut {NoStop}%
\bibitem [{\citenamefont {Hong}\ \emph {et~al.}(2018)\citenamefont {Hong},
  \citenamefont {Ojambati}, \citenamefont {Lagendijk}, \citenamefont {Mosk},\
  and\ \citenamefont {Vos}}]{Hong2018Optica}%
  \BibitemOpen
  \bibfield  {author} {\bibinfo {author} {\bibfnamefont {P.}~\bibnamefont
  {Hong}}, \bibinfo {author} {\bibfnamefont {O.~S.}\ \bibnamefont {Ojambati}},
  \bibinfo {author} {\bibfnamefont {A.}~\bibnamefont {Lagendijk}}, \bibinfo
  {author} {\bibfnamefont {A.~P.}\ \bibnamefont {Mosk}},\ and\ \bibinfo
  {author} {\bibfnamefont {W.~L.}\ \bibnamefont {Vos}},\ }\bibfield  {title}
  {\bibinfo {title} {Three-dimensional spatially resolved optical energy
  density enhanced by wavefront shaping},\ }\href@noop {} {\bibfield  {journal}
  {\bibinfo  {journal} {Optica}\ }\textbf {\bibinfo {volume} {5}},\ \bibinfo
  {pages} {844} (\bibinfo {year} {2018})}\BibitemShut {NoStop}%
\bibitem [{\citenamefont {Uppu}\ \emph {et~al.}(2021)\citenamefont {Uppu},
  \citenamefont {Adhikary}, \citenamefont {Harteveld},\ and\ \citenamefont
  {Vos}}]{Uppu2021PRL}%
  \BibitemOpen
  \bibfield  {author} {\bibinfo {author} {\bibfnamefont {R.}~\bibnamefont
  {Uppu}}, \bibinfo {author} {\bibfnamefont {M.}~\bibnamefont {Adhikary}},
  \bibinfo {author} {\bibfnamefont {C.~A.}\ \bibnamefont {Harteveld}},\ and\
  \bibinfo {author} {\bibfnamefont {W.~L.}\ \bibnamefont {Vos}},\ }\bibfield
  {title} {\bibinfo {title} {Spatially shaping waves to penetrate deep inside a
  forbidden gap},\ }\href@noop {} {\bibfield  {journal} {\bibinfo  {journal}
  {Phys. Rev. Lett.}\ }\textbf {\bibinfo {volume} {126}},\ \bibinfo {pages}
  {177402} (\bibinfo {year} {2021})}\BibitemShut {NoStop}%
\bibitem [{\citenamefont {Bachelard}\ \emph {et~al.}(2012)\citenamefont
  {Bachelard}, \citenamefont {Andreasen}, \citenamefont {Gigan},\ and\
  \citenamefont {Sebbah}}]{bachelard2012taming}%
  \BibitemOpen
  \bibfield  {author} {\bibinfo {author} {\bibfnamefont {N.}~\bibnamefont
  {Bachelard}}, \bibinfo {author} {\bibfnamefont {J.}~\bibnamefont
  {Andreasen}}, \bibinfo {author} {\bibfnamefont {S.}~\bibnamefont {Gigan}},\
  and\ \bibinfo {author} {\bibfnamefont {P.}~\bibnamefont {Sebbah}},\
  }\bibfield  {title} {\bibinfo {title} {Taming random lasers through active
  spatial control of the pump},\ }\href@noop {} {\bibfield  {journal} {\bibinfo
   {journal} {Phys. Rev. Lett.}\ }\textbf {\bibinfo {volume} {109}},\ \bibinfo
  {pages} {033903} (\bibinfo {year} {2012})}\BibitemShut {NoStop}%
\bibitem [{\citenamefont {Qiao}\ \emph {et~al.}(2017)\citenamefont {Qiao},
  \citenamefont {Peng}, \citenamefont {Zheng}, \citenamefont {Ye},\ and\
  \citenamefont {Chen}}]{qiao2017second}%
  \BibitemOpen
  \bibfield  {author} {\bibinfo {author} {\bibfnamefont {Y.}~\bibnamefont
  {Qiao}}, \bibinfo {author} {\bibfnamefont {Y.}~\bibnamefont {Peng}}, \bibinfo
  {author} {\bibfnamefont {Y.}~\bibnamefont {Zheng}}, \bibinfo {author}
  {\bibfnamefont {F.}~\bibnamefont {Ye}},\ and\ \bibinfo {author}
  {\bibfnamefont {X.}~\bibnamefont {Chen}},\ }\bibfield  {title} {\bibinfo
  {title} {Second-harmonic focusing by a nonlinear turbid medium via
  feedback-based wavefront shaping},\ }\href@noop {} {\bibfield  {journal}
  {\bibinfo  {journal} {Opt. Lett.}\ }\textbf {\bibinfo {volume} {42}},\
  \bibinfo {pages} {1895} (\bibinfo {year} {2017})}\BibitemShut {NoStop}%
\bibitem [{\citenamefont {Lib}\ \emph {et~al.}(2020)\citenamefont {Lib},
  \citenamefont {Hasson},\ and\ \citenamefont {Bromberg}}]{lib2020real}%
  \BibitemOpen
  \bibfield  {author} {\bibinfo {author} {\bibfnamefont {O.}~\bibnamefont
  {Lib}}, \bibinfo {author} {\bibfnamefont {G.}~\bibnamefont {Hasson}},\ and\
  \bibinfo {author} {\bibfnamefont {Y.}~\bibnamefont {Bromberg}},\ }\bibfield
  {title} {\bibinfo {title} {Real-time shaping of entangled photons by
  classical control and feedback},\ }\href@noop {} {\bibfield  {journal}
  {\bibinfo  {journal} {Sci. Adv.}\ }\textbf {\bibinfo {volume} {6}},\ \bibinfo
  {pages} {eabb6298} (\bibinfo {year} {2020})}\BibitemShut {NoStop}%
\bibitem [{\citenamefont {Goorden}\ \emph {et~al.}(2014)\citenamefont
  {Goorden}, \citenamefont {Horstmann}, \citenamefont {Mosk}, \citenamefont
  {{\v{S}}kori{\'c}},\ and\ \citenamefont {Pinkse}}]{Goorden2014Optica}%
  \BibitemOpen
  \bibfield  {author} {\bibinfo {author} {\bibfnamefont {S.~A.}\ \bibnamefont
  {Goorden}}, \bibinfo {author} {\bibfnamefont {M.}~\bibnamefont {Horstmann}},
  \bibinfo {author} {\bibfnamefont {A.~P.}\ \bibnamefont {Mosk}}, \bibinfo
  {author} {\bibfnamefont {B.}~\bibnamefont {{\v{S}}kori{\'c}}},\ and\ \bibinfo
  {author} {\bibfnamefont {P.~W.}\ \bibnamefont {Pinkse}},\ }\bibfield  {title}
  {\bibinfo {title} {Quantum-secure authentication of a physical unclonable
  key},\ }\href@noop {} {\bibfield  {journal} {\bibinfo  {journal} {Optica}\
  }\textbf {\bibinfo {volume} {1}},\ \bibinfo {pages} {421} (\bibinfo {year}
  {2014})}\BibitemShut {NoStop}%
\bibitem [{\citenamefont {Wolterink}\ \emph {et~al.}(2016)\citenamefont
  {Wolterink}, \citenamefont {Uppu}, \citenamefont {Ctistis}, \citenamefont
  {Vos}, \citenamefont {Boller},\ and\ \citenamefont
  {Pinkse}}]{wolterink2016programmable}%
  \BibitemOpen
  \bibfield  {author} {\bibinfo {author} {\bibfnamefont {T.~A.}\ \bibnamefont
  {Wolterink}}, \bibinfo {author} {\bibfnamefont {R.}~\bibnamefont {Uppu}},
  \bibinfo {author} {\bibfnamefont {G.}~\bibnamefont {Ctistis}}, \bibinfo
  {author} {\bibfnamefont {W.~L.}\ \bibnamefont {Vos}}, \bibinfo {author}
  {\bibfnamefont {K.-J.}\ \bibnamefont {Boller}},\ and\ \bibinfo {author}
  {\bibfnamefont {P.~W.}\ \bibnamefont {Pinkse}},\ }\bibfield  {title}
  {\bibinfo {title} {Programmable two-photon quantum interference in 10 3
  channels in opaque scattering media},\ }\href@noop {} {\bibfield  {journal}
  {\bibinfo  {journal} {Phys. Rev. A}\ }\textbf {\bibinfo {volume} {93}},\
  \bibinfo {pages} {053817} (\bibinfo {year} {2016})}\BibitemShut {NoStop}%
\bibitem [{\citenamefont {Valencia}\ \emph {et~al.}(2020)\citenamefont
  {Valencia}, \citenamefont {Goel}, \citenamefont {McCutcheon}, \citenamefont
  {Defienne},\ and\ \citenamefont {Malik}}]{valencia2020unscrambling}%
  \BibitemOpen
  \bibfield  {author} {\bibinfo {author} {\bibfnamefont {N.~H.}\ \bibnamefont
  {Valencia}}, \bibinfo {author} {\bibfnamefont {S.}~\bibnamefont {Goel}},
  \bibinfo {author} {\bibfnamefont {W.}~\bibnamefont {McCutcheon}}, \bibinfo
  {author} {\bibfnamefont {H.}~\bibnamefont {Defienne}},\ and\ \bibinfo
  {author} {\bibfnamefont {M.}~\bibnamefont {Malik}},\ }\bibfield  {title}
  {\bibinfo {title} {Unscrambling entanglement through a complex medium},\
  }\href@noop {} {\bibfield  {journal} {\bibinfo  {journal} {Nat. Phys.}\
  }\textbf {\bibinfo {volume} {16}},\ \bibinfo {pages} {1112} (\bibinfo {year}
  {2020})}\BibitemShut {NoStop}%
\bibitem [{\citenamefont {Leedumrongwatthanakun}\ \emph
  {et~al.}(2020)\citenamefont {Leedumrongwatthanakun}, \citenamefont
  {Innocenti}, \citenamefont {Defienne}, \citenamefont {Juffmann},
  \citenamefont {Ferraro}, \citenamefont {Paternostro},\ and\ \citenamefont
  {Gigan}}]{leedumrongwatthanakun2020programmable}%
  \BibitemOpen
  \bibfield  {author} {\bibinfo {author} {\bibfnamefont {S.}~\bibnamefont
  {Leedumrongwatthanakun}}, \bibinfo {author} {\bibfnamefont {L.}~\bibnamefont
  {Innocenti}}, \bibinfo {author} {\bibfnamefont {H.}~\bibnamefont {Defienne}},
  \bibinfo {author} {\bibfnamefont {T.}~\bibnamefont {Juffmann}}, \bibinfo
  {author} {\bibfnamefont {A.}~\bibnamefont {Ferraro}}, \bibinfo {author}
  {\bibfnamefont {M.}~\bibnamefont {Paternostro}},\ and\ \bibinfo {author}
  {\bibfnamefont {S.}~\bibnamefont {Gigan}},\ }\bibfield  {title} {\bibinfo
  {title} {Programmable linear quantum networks with a multimode fibre},\
  }\href@noop {} {\bibfield  {journal} {\bibinfo  {journal} {Nat. Photon.}\
  }\textbf {\bibinfo {volume} {14}},\ \bibinfo {pages} {139} (\bibinfo {year}
  {2020})}\BibitemShut {NoStop}%
\bibitem [{\citenamefont {Kuznetsov}\ \emph {et~al.}(2016)\citenamefont
  {Kuznetsov}, \citenamefont {Miroshnichenko}, \citenamefont {Brongersma},
  \citenamefont {Kivshar},\ and\ \citenamefont
  {Luk’yanchuk}}]{kuznetsov2016optically}%
  \BibitemOpen
  \bibfield  {author} {\bibinfo {author} {\bibfnamefont {A.~I.}\ \bibnamefont
  {Kuznetsov}}, \bibinfo {author} {\bibfnamefont {A.~E.}\ \bibnamefont
  {Miroshnichenko}}, \bibinfo {author} {\bibfnamefont {M.~L.}\ \bibnamefont
  {Brongersma}}, \bibinfo {author} {\bibfnamefont {Y.~S.}\ \bibnamefont
  {Kivshar}},\ and\ \bibinfo {author} {\bibfnamefont {B.}~\bibnamefont
  {Luk’yanchuk}},\ }\bibfield  {title} {\bibinfo {title} {Optically resonant
  dielectric nanostructures},\ }\href@noop {} {\bibfield  {journal} {\bibinfo
  {journal} {Science}\ }\textbf {\bibinfo {volume} {354}} (\bibinfo {year}
  {2016})}\BibitemShut {NoStop}%
\bibitem [{\citenamefont {Staude}\ and\ \citenamefont
  {Schilling}(2017)}]{staude2017metamaterial}%
  \BibitemOpen
  \bibfield  {author} {\bibinfo {author} {\bibfnamefont {I.}~\bibnamefont
  {Staude}}\ and\ \bibinfo {author} {\bibfnamefont {J.}~\bibnamefont
  {Schilling}},\ }\bibfield  {title} {\bibinfo {title} {Metamaterial-inspired
  silicon nanophotonics},\ }\href@noop {} {\bibfield  {journal} {\bibinfo
  {journal} {Nat. Photon.}\ }\textbf {\bibinfo {volume} {11}},\ \bibinfo
  {pages} {274} (\bibinfo {year} {2017})}\BibitemShut {NoStop}%
\bibitem [{\citenamefont {Geffrin}\ \emph {et~al.}(2012)\citenamefont
  {Geffrin}, \citenamefont {Garc{\'\i}a-C{\'a}mara}, \citenamefont
  {G{\'o}mez-Medina}, \citenamefont {Albella}, \citenamefont
  {Froufe-P{\'e}rez}, \citenamefont {Eyraud}, \citenamefont {Litman},
  \citenamefont {Vaillon}, \citenamefont {Gonz{\'a}lez}, \citenamefont
  {Nieto-Vesperinas} \emph {et~al.}}]{geffrin2012magnetic}%
  \BibitemOpen
  \bibfield  {author} {\bibinfo {author} {\bibfnamefont {J.-M.}\ \bibnamefont
  {Geffrin}}, \bibinfo {author} {\bibfnamefont {B.}~\bibnamefont
  {Garc{\'\i}a-C{\'a}mara}}, \bibinfo {author} {\bibfnamefont {R.}~\bibnamefont
  {G{\'o}mez-Medina}}, \bibinfo {author} {\bibfnamefont {P.}~\bibnamefont
  {Albella}}, \bibinfo {author} {\bibfnamefont {L.}~\bibnamefont
  {Froufe-P{\'e}rez}}, \bibinfo {author} {\bibfnamefont {C.}~\bibnamefont
  {Eyraud}}, \bibinfo {author} {\bibfnamefont {A.}~\bibnamefont {Litman}},
  \bibinfo {author} {\bibfnamefont {R.}~\bibnamefont {Vaillon}}, \bibinfo
  {author} {\bibfnamefont {F.}~\bibnamefont {Gonz{\'a}lez}}, \bibinfo {author}
  {\bibfnamefont {M.}~\bibnamefont {Nieto-Vesperinas}}, \emph {et~al.},\
  }\bibfield  {title} {\bibinfo {title} {Magnetic and electric coherence in
  forward-and back-scattered electromagnetic waves by a single dielectric
  subwavelength sphere},\ }\href@noop {} {\bibfield  {journal} {\bibinfo
  {journal} {Nat. Commun.}\ }\textbf {\bibinfo {volume} {3}},\ \bibinfo {pages}
  {1} (\bibinfo {year} {2012})}\BibitemShut {NoStop}%
\bibitem [{\citenamefont {Miroshnichenko}\ \emph {et~al.}(2015)\citenamefont
  {Miroshnichenko}, \citenamefont {Evlyukhin}, \citenamefont {Yu},
  \citenamefont {Bakker}, \citenamefont {Chipouline}, \citenamefont
  {Kuznetsov}, \citenamefont {Luk’yanchuk}, \citenamefont {Chichkov},\ and\
  \citenamefont {Kivshar}}]{miroshnichenko2015nonradiating}%
  \BibitemOpen
  \bibfield  {author} {\bibinfo {author} {\bibfnamefont {A.~E.}\ \bibnamefont
  {Miroshnichenko}}, \bibinfo {author} {\bibfnamefont {A.~B.}\ \bibnamefont
  {Evlyukhin}}, \bibinfo {author} {\bibfnamefont {Y.~F.}\ \bibnamefont {Yu}},
  \bibinfo {author} {\bibfnamefont {R.~M.}\ \bibnamefont {Bakker}}, \bibinfo
  {author} {\bibfnamefont {A.}~\bibnamefont {Chipouline}}, \bibinfo {author}
  {\bibfnamefont {A.~I.}\ \bibnamefont {Kuznetsov}}, \bibinfo {author}
  {\bibfnamefont {B.}~\bibnamefont {Luk’yanchuk}}, \bibinfo {author}
  {\bibfnamefont {B.~N.}\ \bibnamefont {Chichkov}},\ and\ \bibinfo {author}
  {\bibfnamefont {Y.~S.}\ \bibnamefont {Kivshar}},\ }\bibfield  {title}
  {\bibinfo {title} {Nonradiating anapole modes in dielectric nanoparticles},\
  }\href@noop {} {\bibfield  {journal} {\bibinfo  {journal} {Nat. Commun.}\
  }\textbf {\bibinfo {volume} {6}},\ \bibinfo {pages} {1} (\bibinfo {year}
  {2015})}\BibitemShut {NoStop}%
\bibitem [{\citenamefont {Shamkhi}\ \emph {et~al.}(2019)\citenamefont
  {Shamkhi}, \citenamefont {Baryshnikova}, \citenamefont {Sayanskiy},
  \citenamefont {Kapitanova}, \citenamefont {Terekhov}, \citenamefont {Belov},
  \citenamefont {Karabchevsky}, \citenamefont {Evlyukhin}, \citenamefont
  {Kivshar},\ and\ \citenamefont {Shalin}}]{shamkhi2019transverse}%
  \BibitemOpen
  \bibfield  {author} {\bibinfo {author} {\bibfnamefont {H.~K.}\ \bibnamefont
  {Shamkhi}}, \bibinfo {author} {\bibfnamefont {K.~V.}\ \bibnamefont
  {Baryshnikova}}, \bibinfo {author} {\bibfnamefont {A.}~\bibnamefont
  {Sayanskiy}}, \bibinfo {author} {\bibfnamefont {P.}~\bibnamefont
  {Kapitanova}}, \bibinfo {author} {\bibfnamefont {P.~D.}\ \bibnamefont
  {Terekhov}}, \bibinfo {author} {\bibfnamefont {P.}~\bibnamefont {Belov}},
  \bibinfo {author} {\bibfnamefont {A.}~\bibnamefont {Karabchevsky}}, \bibinfo
  {author} {\bibfnamefont {A.~B.}\ \bibnamefont {Evlyukhin}}, \bibinfo {author}
  {\bibfnamefont {Y.}~\bibnamefont {Kivshar}},\ and\ \bibinfo {author}
  {\bibfnamefont {A.~S.}\ \bibnamefont {Shalin}},\ }\bibfield  {title}
  {\bibinfo {title} {Transverse scattering and generalized kerker effects in
  all-dielectric mie-resonant metaoptics},\ }\href@noop {} {\bibfield
  {journal} {\bibinfo  {journal} {Phys. Rev. Lett.}\ }\textbf {\bibinfo
  {volume} {122}},\ \bibinfo {pages} {193905} (\bibinfo {year}
  {2019})}\BibitemShut {NoStop}%
\bibitem [{\citenamefont {Rutckaia}\ \emph {et~al.}(2017)\citenamefont
  {Rutckaia}, \citenamefont {Heyroth}, \citenamefont {Novikov}, \citenamefont
  {Shaleev}, \citenamefont {Petrov},\ and\ \citenamefont
  {Schilling}}]{rutckaia2017quantum}%
  \BibitemOpen
  \bibfield  {author} {\bibinfo {author} {\bibfnamefont {V.}~\bibnamefont
  {Rutckaia}}, \bibinfo {author} {\bibfnamefont {F.}~\bibnamefont {Heyroth}},
  \bibinfo {author} {\bibfnamefont {A.}~\bibnamefont {Novikov}}, \bibinfo
  {author} {\bibfnamefont {M.}~\bibnamefont {Shaleev}}, \bibinfo {author}
  {\bibfnamefont {M.}~\bibnamefont {Petrov}},\ and\ \bibinfo {author}
  {\bibfnamefont {J.}~\bibnamefont {Schilling}},\ }\bibfield  {title} {\bibinfo
  {title} {Quantum dot emission driven by mie resonances in silicon
  nanostructures},\ }\href@noop {} {\bibfield  {journal} {\bibinfo  {journal}
  {Nano Lett.}\ }\textbf {\bibinfo {volume} {17}},\ \bibinfo {pages} {6886}
  (\bibinfo {year} {2017})}\BibitemShut {NoStop}%
\bibitem [{\citenamefont {Cihan}\ \emph {et~al.}(2018)\citenamefont {Cihan},
  \citenamefont {Curto}, \citenamefont {Raza}, \citenamefont {Kik},\ and\
  \citenamefont {Brongersma}}]{cihan2018silicon}%
  \BibitemOpen
  \bibfield  {author} {\bibinfo {author} {\bibfnamefont {A.~F.}\ \bibnamefont
  {Cihan}}, \bibinfo {author} {\bibfnamefont {A.~G.}\ \bibnamefont {Curto}},
  \bibinfo {author} {\bibfnamefont {S.}~\bibnamefont {Raza}}, \bibinfo {author}
  {\bibfnamefont {P.~G.}\ \bibnamefont {Kik}},\ and\ \bibinfo {author}
  {\bibfnamefont {M.~L.}\ \bibnamefont {Brongersma}},\ }\bibfield  {title}
  {\bibinfo {title} {Silicon mie resonators for highly directional light
  emission from monolayer mos 2},\ }\href@noop {} {\bibfield  {journal}
  {\bibinfo  {journal} {Nat. Photon.}\ }\textbf {\bibinfo {volume} {12}},\
  \bibinfo {pages} {284} (\bibinfo {year} {2018})}\BibitemShut {NoStop}%
\bibitem [{\citenamefont {Gongora}\ \emph {et~al.}(2017)\citenamefont
  {Gongora}, \citenamefont {Miroshnichenko}, \citenamefont {Kivshar},\ and\
  \citenamefont {Fratalocchi}}]{gongora2017anapole}%
  \BibitemOpen
  \bibfield  {author} {\bibinfo {author} {\bibfnamefont {J.~S.~T.}\
  \bibnamefont {Gongora}}, \bibinfo {author} {\bibfnamefont {A.~E.}\
  \bibnamefont {Miroshnichenko}}, \bibinfo {author} {\bibfnamefont {Y.~S.}\
  \bibnamefont {Kivshar}},\ and\ \bibinfo {author} {\bibfnamefont
  {A.}~\bibnamefont {Fratalocchi}},\ }\bibfield  {title} {\bibinfo {title}
  {Anapole nanolasers for mode-locking and ultrafast pulse generation},\
  }\href@noop {} {\bibfield  {journal} {\bibinfo  {journal} {Nat. Commun.}\
  }\textbf {\bibinfo {volume} {8}},\ \bibinfo {pages} {1} (\bibinfo {year}
  {2017})}\BibitemShut {NoStop}%
\bibitem [{\citenamefont {Tiguntseva}\ \emph {et~al.}(2020)\citenamefont
  {Tiguntseva}, \citenamefont {Koshelev}, \citenamefont {Furasova},
  \citenamefont {Tonkaev}, \citenamefont {Mikhailovskii}, \citenamefont
  {Ushakova}, \citenamefont {Baranov}, \citenamefont {Shegai}, \citenamefont
  {Zakhidov}, \citenamefont {Kivshar} \emph {et~al.}}]{tiguntseva2020room}%
  \BibitemOpen
  \bibfield  {author} {\bibinfo {author} {\bibfnamefont {E.}~\bibnamefont
  {Tiguntseva}}, \bibinfo {author} {\bibfnamefont {K.}~\bibnamefont
  {Koshelev}}, \bibinfo {author} {\bibfnamefont {A.}~\bibnamefont {Furasova}},
  \bibinfo {author} {\bibfnamefont {P.}~\bibnamefont {Tonkaev}}, \bibinfo
  {author} {\bibfnamefont {V.}~\bibnamefont {Mikhailovskii}}, \bibinfo {author}
  {\bibfnamefont {E.~V.}\ \bibnamefont {Ushakova}}, \bibinfo {author}
  {\bibfnamefont {D.~G.}\ \bibnamefont {Baranov}}, \bibinfo {author}
  {\bibfnamefont {T.}~\bibnamefont {Shegai}}, \bibinfo {author} {\bibfnamefont
  {A.~A.}\ \bibnamefont {Zakhidov}}, \bibinfo {author} {\bibfnamefont
  {Y.}~\bibnamefont {Kivshar}}, \emph {et~al.},\ }\bibfield  {title} {\bibinfo
  {title} {Room-temperature lasing from mie-resonant nonplasmonic
  nanoparticles},\ }\href@noop {} {\bibfield  {journal} {\bibinfo  {journal}
  {ACS nano}\ }\textbf {\bibinfo {volume} {14}},\ \bibinfo {pages} {8149}
  (\bibinfo {year} {2020})}\BibitemShut {NoStop}%
\bibitem [{\citenamefont {Grinblat}\ \emph {et~al.}(2016)\citenamefont
  {Grinblat}, \citenamefont {Li}, \citenamefont {Nielsen}, \citenamefont
  {Oulton},\ and\ \citenamefont {Maier}}]{grinblat2016enhanced}%
  \BibitemOpen
  \bibfield  {author} {\bibinfo {author} {\bibfnamefont {G.}~\bibnamefont
  {Grinblat}}, \bibinfo {author} {\bibfnamefont {Y.}~\bibnamefont {Li}},
  \bibinfo {author} {\bibfnamefont {M.~P.}\ \bibnamefont {Nielsen}}, \bibinfo
  {author} {\bibfnamefont {R.~F.}\ \bibnamefont {Oulton}},\ and\ \bibinfo
  {author} {\bibfnamefont {S.~A.}\ \bibnamefont {Maier}},\ }\bibfield  {title}
  {\bibinfo {title} {Enhanced third harmonic generation in single germanium
  nanodisks excited at the anapole mode},\ }\href@noop {} {\bibfield  {journal}
  {\bibinfo  {journal} {Nano Lett.}\ }\textbf {\bibinfo {volume} {16}},\
  \bibinfo {pages} {4635} (\bibinfo {year} {2016})}\BibitemShut {NoStop}%
\bibitem [{\citenamefont {Xu}\ \emph {et~al.}(2018)\citenamefont {Xu},
  \citenamefont {Rahmani}, \citenamefont {Kamali}, \citenamefont
  {Lamprianidis}, \citenamefont {Ghirardini}, \citenamefont {Sautter},
  \citenamefont {Camacho-Morales}, \citenamefont {Chen}, \citenamefont {Parry},
  \citenamefont {Staude} \emph {et~al.}}]{xu2018boosting}%
  \BibitemOpen
  \bibfield  {author} {\bibinfo {author} {\bibfnamefont {L.}~\bibnamefont
  {Xu}}, \bibinfo {author} {\bibfnamefont {M.}~\bibnamefont {Rahmani}},
  \bibinfo {author} {\bibfnamefont {K.~Z.}\ \bibnamefont {Kamali}}, \bibinfo
  {author} {\bibfnamefont {A.}~\bibnamefont {Lamprianidis}}, \bibinfo {author}
  {\bibfnamefont {L.}~\bibnamefont {Ghirardini}}, \bibinfo {author}
  {\bibfnamefont {J.}~\bibnamefont {Sautter}}, \bibinfo {author} {\bibfnamefont
  {R.}~\bibnamefont {Camacho-Morales}}, \bibinfo {author} {\bibfnamefont
  {H.}~\bibnamefont {Chen}}, \bibinfo {author} {\bibfnamefont {M.}~\bibnamefont
  {Parry}}, \bibinfo {author} {\bibfnamefont {I.}~\bibnamefont {Staude}}, \emph
  {et~al.},\ }\bibfield  {title} {\bibinfo {title} {Boosting third-harmonic
  generation by a mirror-enhanced anapole resonator},\ }\href@noop {}
  {\bibfield  {journal} {\bibinfo  {journal} {Light Sci. Appl.}\ }\textbf
  {\bibinfo {volume} {7}},\ \bibinfo {pages} {1} (\bibinfo {year}
  {2018})}\BibitemShut {NoStop}%
\bibitem [{\citenamefont {Wei}\ \emph {et~al.}(2016)\citenamefont {Wei},
  \citenamefont {Xi}, \citenamefont {Bhattacharya},\ and\ \citenamefont
  {Urbach}}]{wei2016excitation}%
  \BibitemOpen
  \bibfield  {author} {\bibinfo {author} {\bibfnamefont {L.}~\bibnamefont
  {Wei}}, \bibinfo {author} {\bibfnamefont {Z.}~\bibnamefont {Xi}}, \bibinfo
  {author} {\bibfnamefont {N.}~\bibnamefont {Bhattacharya}},\ and\ \bibinfo
  {author} {\bibfnamefont {H.~P.}\ \bibnamefont {Urbach}},\ }\bibfield  {title}
  {\bibinfo {title} {Excitation of the radiationless anapole mode},\
  }\href@noop {} {\bibfield  {journal} {\bibinfo  {journal} {Optica}\ }\textbf
  {\bibinfo {volume} {3}},\ \bibinfo {pages} {799} (\bibinfo {year}
  {2016})}\BibitemShut {NoStop}%
\bibitem [{\citenamefont {Raybould}\ \emph {et~al.}(2017)\citenamefont
  {Raybould}, \citenamefont {Fedotov}, \citenamefont {Papasimakis},
  \citenamefont {Youngs},\ and\ \citenamefont
  {Zheludev}}]{raybould2017exciting}%
  \BibitemOpen
  \bibfield  {author} {\bibinfo {author} {\bibfnamefont {T.}~\bibnamefont
  {Raybould}}, \bibinfo {author} {\bibfnamefont {V.~A.}\ \bibnamefont
  {Fedotov}}, \bibinfo {author} {\bibfnamefont {N.}~\bibnamefont
  {Papasimakis}}, \bibinfo {author} {\bibfnamefont {I.}~\bibnamefont
  {Youngs}},\ and\ \bibinfo {author} {\bibfnamefont {N.~I.}\ \bibnamefont
  {Zheludev}},\ }\bibfield  {title} {\bibinfo {title} {Exciting dynamic
  anapoles with electromagnetic doughnut pulses},\ }\href@noop {} {\bibfield
  {journal} {\bibinfo  {journal} {Appl. Phys. Lett.}\ }\textbf {\bibinfo
  {volume} {111}},\ \bibinfo {pages} {081104} (\bibinfo {year}
  {2017})}\BibitemShut {NoStop}%
\bibitem [{\citenamefont {Saadabad}\ \emph {et~al.}(2021)\citenamefont
  {Saadabad}, \citenamefont {Cai}, \citenamefont {Deng}, \citenamefont {Xu},\
  and\ \citenamefont {Miroshnichenko}}]{saadabad2021structured}%
  \BibitemOpen
  \bibfield  {author} {\bibinfo {author} {\bibfnamefont {R.~M.}\ \bibnamefont
  {Saadabad}}, \bibinfo {author} {\bibfnamefont {M.}~\bibnamefont {Cai}},
  \bibinfo {author} {\bibfnamefont {F.}~\bibnamefont {Deng}}, \bibinfo {author}
  {\bibfnamefont {L.}~\bibnamefont {Xu}},\ and\ \bibinfo {author}
  {\bibfnamefont {A.~E.}\ \bibnamefont {Miroshnichenko}},\ }\bibfield  {title}
  {\bibinfo {title} {Structured light excitation of toroidal dipoles in
  dielectric nanodisks},\ }\href@noop {} {\bibfield  {journal} {\bibinfo
  {journal} {Phys. Rev. B}\ }\textbf {\bibinfo {volume} {104}},\ \bibinfo
  {pages} {165402} (\bibinfo {year} {2021})}\BibitemShut {NoStop}%
\bibitem [{\citenamefont {Rotter}\ and\ \citenamefont
  {Gigan}(2017)}]{Rotter2017RMP}%
  \BibitemOpen
  \bibfield  {author} {\bibinfo {author} {\bibfnamefont {S.}~\bibnamefont
  {Rotter}}\ and\ \bibinfo {author} {\bibfnamefont {S.}~\bibnamefont {Gigan}},\
  }\bibfield  {title} {\bibinfo {title} {Light fields in complex media:
  Mesoscopic scattering meets wave control},\ }\href
  {https://doi.org/10.1103/RevModPhys.89.015005} {\bibfield  {journal}
  {\bibinfo  {journal} {Rev. Mod. Phys.}\ }\textbf {\bibinfo {volume} {89}},\
  \bibinfo {pages} {015005} (\bibinfo {year} {2017})}\BibitemShut {NoStop}%
\bibitem [{Note1()}]{Note1}%
  \BibitemOpen
  \bibinfo {note} {The transition matrix, that originally originates in
  high-energy physics should not to be confused with the transmission matrix
  that originates from mesoscopic physics.}\BibitemShut {Stop}%
\bibitem [{\citenamefont {Lagendijk}\ and\ \citenamefont {van
  Tiggelen}(1996)}]{Lagendijk1996PhysRep}%
  \BibitemOpen
  \bibfield  {author} {\bibinfo {author} {\bibfnamefont {A.}~\bibnamefont
  {Lagendijk}}\ and\ \bibinfo {author} {\bibfnamefont {B.~A.}\ \bibnamefont
  {van Tiggelen}},\ }\bibfield  {title} {\bibinfo {title} {Resonant multiple
  scattering of light},\ }\href {https://doi.org/10.1016/0370-1573(95)00065-8}
  {\bibfield  {journal} {\bibinfo  {journal} {Phys. Rep.}\ }\textbf {\bibinfo
  {volume} {270}},\ \bibinfo {pages} {143} (\bibinfo {year}
  {1996})}\BibitemShut {NoStop}%
\bibitem [{\citenamefont {van Rossum}\ and\ \citenamefont
  {Nieuwenhuizen}(1999)}]{vanRossum1999RMP}%
  \BibitemOpen
  \bibfield  {author} {\bibinfo {author} {\bibfnamefont {M.~C.~W.}\
  \bibnamefont {van Rossum}}\ and\ \bibinfo {author} {\bibfnamefont {T.~M.}\
  \bibnamefont {Nieuwenhuizen}},\ }\bibfield  {title} {\bibinfo {title}
  {Multiple scattering of classical waves: microscopy, mesoscopy, and
  diffusion},\ }\href {https://doi.org/10.1103/RevModPhys.71.313} {\bibfield
  {journal} {\bibinfo  {journal} {Rev. Mod. Phys.}\ }\textbf {\bibinfo {volume}
  {71}},\ \bibinfo {pages} {313} (\bibinfo {year} {1999})}\BibitemShut
  {NoStop}%
\bibitem [{Note2()}]{Note2}%
  \BibitemOpen
  \bibinfo {note} {Thus our study is complementary to the recently discovered
  scattering phenomena of mutual extinction and transparency, where one
  explicitly considers the interference between scattered waves and the
  incident wave~\cite {lagendijk2020mutual}.}\BibitemShut {Stop}%
\bibitem [{\citenamefont {Vellekoop}\ and\ \citenamefont
  {Mosk}(2008)}]{vellekoop2008universal}%
  \BibitemOpen
  \bibfield  {author} {\bibinfo {author} {\bibfnamefont {I.~M.}\ \bibnamefont
  {Vellekoop}}\ and\ \bibinfo {author} {\bibfnamefont {A.}~\bibnamefont
  {Mosk}},\ }\bibfield  {title} {\bibinfo {title} {Universal optimal
  transmission of light through disordered materials},\ }\href@noop {}
  {\bibfield  {journal} {\bibinfo  {journal} {Phys. Rev. Lett.}\ }\textbf
  {\bibinfo {volume} {101}},\ \bibinfo {pages} {120601} (\bibinfo {year}
  {2008})}\BibitemShut {NoStop}%
\bibitem [{\citenamefont {Yu}\ \emph {et~al.}(2013)\citenamefont {Yu},
  \citenamefont {Hillman}, \citenamefont {Choi}, \citenamefont {Lee},
  \citenamefont {Feld}, \citenamefont {Dasari},\ and\ \citenamefont
  {Park}}]{yu2013measuring}%
  \BibitemOpen
  \bibfield  {author} {\bibinfo {author} {\bibfnamefont {H.}~\bibnamefont
  {Yu}}, \bibinfo {author} {\bibfnamefont {T.~R.}\ \bibnamefont {Hillman}},
  \bibinfo {author} {\bibfnamefont {W.}~\bibnamefont {Choi}}, \bibinfo {author}
  {\bibfnamefont {J.~O.}\ \bibnamefont {Lee}}, \bibinfo {author} {\bibfnamefont
  {M.~S.}\ \bibnamefont {Feld}}, \bibinfo {author} {\bibfnamefont {R.~R.}\
  \bibnamefont {Dasari}},\ and\ \bibinfo {author} {\bibfnamefont
  {Y.}~\bibnamefont {Park}},\ }\bibfield  {title} {\bibinfo {title} {Measuring
  large optical transmission matrices of disordered media},\ }\href@noop {}
  {\bibfield  {journal} {\bibinfo  {journal} {Phys. Rev. Lett.}\ }\textbf
  {\bibinfo {volume} {111}},\ \bibinfo {pages} {153902} (\bibinfo {year}
  {2013})}\BibitemShut {NoStop}%
\bibitem [{\citenamefont {Goetschy}\ and\ \citenamefont
  {Stone}(2013)}]{goetschy2013filtering}%
  \BibitemOpen
  \bibfield  {author} {\bibinfo {author} {\bibfnamefont {A.}~\bibnamefont
  {Goetschy}}\ and\ \bibinfo {author} {\bibfnamefont {A.}~\bibnamefont
  {Stone}},\ }\bibfield  {title} {\bibinfo {title} {Filtering random matrices:
  the effect of incomplete channel control in multiple scattering},\
  }\href@noop {} {\bibfield  {journal} {\bibinfo  {journal} {Phys. Rev. Lett.}\
  }\textbf {\bibinfo {volume} {111}},\ \bibinfo {pages} {063901} (\bibinfo
  {year} {2013})}\BibitemShut {NoStop}%
\bibitem [{\citenamefont {Lagendijk}\ \emph {et~al.}(2020)\citenamefont
  {Lagendijk}, \citenamefont {Mosk},\ and\ \citenamefont
  {Vos}}]{lagendijk2020mutual}%
  \BibitemOpen
  \bibfield  {author} {\bibinfo {author} {\bibfnamefont {A.}~\bibnamefont
  {Lagendijk}}, \bibinfo {author} {\bibfnamefont {A.~P.}\ \bibnamefont
  {Mosk}},\ and\ \bibinfo {author} {\bibfnamefont {W.~L.}\ \bibnamefont
  {Vos}},\ }\bibfield  {title} {\bibinfo {title} {Mutual extinction and
  transparency of multiple incident light waves},\ }\href@noop {} {\bibfield
  {journal} {\bibinfo  {journal} {Europhys. Lett.}\ }\textbf {\bibinfo {volume}
  {130}},\ \bibinfo {pages} {34002} (\bibinfo {year} {2020})}\BibitemShut
  {NoStop}%
\bibitem [{\citenamefont {Rates}\ \emph {et~al.}(2021)\citenamefont {Rates},
  \citenamefont {Lagendijk}, \citenamefont {Akdemir}, \citenamefont {Mosk},\
  and\ \citenamefont {Vos}}]{rates2021observation}%
  \BibitemOpen
  \bibfield  {author} {\bibinfo {author} {\bibfnamefont {A.}~\bibnamefont
  {Rates}}, \bibinfo {author} {\bibfnamefont {A.}~\bibnamefont {Lagendijk}},
  \bibinfo {author} {\bibfnamefont {O.}~\bibnamefont {Akdemir}}, \bibinfo
  {author} {\bibfnamefont {A.~P.}\ \bibnamefont {Mosk}},\ and\ \bibinfo
  {author} {\bibfnamefont {W.~L.}\ \bibnamefont {Vos}},\ }\bibfield  {title}
  {\bibinfo {title} {Observation of mutual extinction and transparency in light
  scattering},\ }\href@noop {} {\bibfield  {journal} {\bibinfo  {journal}
  {Phys. Rev. A}\ }\textbf {\bibinfo {volume} {104}},\ \bibinfo {pages}
  {043515} (\bibinfo {year} {2021})}\BibitemShut {NoStop}%
\bibitem [{\citenamefont {William H.~Press}\ and\ \citenamefont
  {Flannery}(2007)}]{Press2007book}%
  \BibitemOpen
  \bibfield  {author} {\bibinfo {author} {\bibfnamefont {W.~T.~V.}\
  \bibnamefont {William H.~Press}, \bibfnamefont {Saul A.~Teukolsky}}\ and\
  \bibinfo {author} {\bibfnamefont {B.~P.}\ \bibnamefont {Flannery}},\
  }\href@noop {} {\emph {\bibinfo {title} {Numerical Recipes}}},\ \bibinfo
  {edition} {3rd}\ ed.\ (\bibinfo  {publisher} {Cambrdige University Press},\
  \bibinfo {year} {2007})\BibitemShut {NoStop}%
\bibitem [{\citenamefont {Miller}(2019)}]{Miller2019AOP}%
  \BibitemOpen
  \bibfield  {author} {\bibinfo {author} {\bibfnamefont {D.~A.}\ \bibnamefont
  {Miller}},\ }\bibfield  {title} {\bibinfo {title} {Waves, modes,
  communications, and optics: a tutorial},\ }\href@noop {} {\bibfield
  {journal} {\bibinfo  {journal} {Adv. Opt. Photonics}\ }\textbf {\bibinfo
  {volume} {11}},\ \bibinfo {pages} {679} (\bibinfo {year} {2019})}\BibitemShut
  {NoStop}%
\bibitem [{\citenamefont {Green}(2008)}]{green2008self}%
  \BibitemOpen
  \bibfield  {author} {\bibinfo {author} {\bibfnamefont {M.~A.}\ \bibnamefont
  {Green}},\ }\bibfield  {title} {\bibinfo {title} {Self-consistent optical
  parameters of intrinsic silicon at 300 k including temperature
  coefficients},\ }\href@noop {} {\bibfield  {journal} {\bibinfo  {journal}
  {Sol. Energy Mater Sol. Cells}\ }\textbf {\bibinfo {volume} {92}},\ \bibinfo
  {pages} {1305} (\bibinfo {year} {2008})}\BibitemShut {NoStop}%
\bibitem [{\citenamefont {Jellison~Jr}(1992)}]{jellison1992optical}%
  \BibitemOpen
  \bibfield  {author} {\bibinfo {author} {\bibfnamefont {G.}~\bibnamefont
  {Jellison~Jr}},\ }\bibfield  {title} {\bibinfo {title} {Optical functions of
  gaas, gap, and ge determined by two-channel polarization modulation
  ellipsometry},\ }\href@noop {} {\bibfield  {journal} {\bibinfo  {journal}
  {Opt. Mater.}\ }\textbf {\bibinfo {volume} {1}},\ \bibinfo {pages} {151}
  (\bibinfo {year} {1992})}\BibitemShut {NoStop}%
\bibitem [{\citenamefont {Bohren}\ and\ \citenamefont
  {Huffman}(2008)}]{Bohren2008Book}%
  \BibitemOpen
  \bibfield  {author} {\bibinfo {author} {\bibfnamefont {C.~F.}\ \bibnamefont
  {Bohren}}\ and\ \bibinfo {author} {\bibfnamefont {D.~R.}\ \bibnamefont
  {Huffman}},\ }\href@noop {} {\emph {\bibinfo {title} {Absorption and
  scattering of light by small particles}}}\ (\bibinfo  {publisher} {John Wiley
  \& Sons},\ \bibinfo {year} {2008})\BibitemShut {NoStop}%
\bibitem [{\citenamefont {Frezza}\ \emph {et~al.}(2018)\citenamefont {Frezza},
  \citenamefont {Mangini},\ and\ \citenamefont {Tedeschi}}]{Frezza2018JOSAA}%
  \BibitemOpen
  \bibfield  {author} {\bibinfo {author} {\bibfnamefont {F.}~\bibnamefont
  {Frezza}}, \bibinfo {author} {\bibfnamefont {F.}~\bibnamefont {Mangini}},\
  and\ \bibinfo {author} {\bibfnamefont {N.}~\bibnamefont {Tedeschi}},\
  }\bibfield  {title} {\bibinfo {title} {Introduction to electromagnetic
  scattering: tutorial},\ }\href@noop {} {\bibfield  {journal} {\bibinfo
  {journal} {JOSA A}\ }\textbf {\bibinfo {volume} {35}},\ \bibinfo {pages}
  {163} (\bibinfo {year} {2018})}\BibitemShut {NoStop}%
\bibitem [{\citenamefont {Aulbach}\ \emph {et~al.}(2011)\citenamefont
  {Aulbach}, \citenamefont {Gjonaj}, \citenamefont {Johnson}, \citenamefont
  {Mosk},\ and\ \citenamefont {Lagendijk}}]{Aulbach2011PRL}%
  \BibitemOpen
  \bibfield  {author} {\bibinfo {author} {\bibfnamefont {J.}~\bibnamefont
  {Aulbach}}, \bibinfo {author} {\bibfnamefont {B.}~\bibnamefont {Gjonaj}},
  \bibinfo {author} {\bibfnamefont {P.~M.}\ \bibnamefont {Johnson}}, \bibinfo
  {author} {\bibfnamefont {A.~P.}\ \bibnamefont {Mosk}},\ and\ \bibinfo
  {author} {\bibfnamefont {A.}~\bibnamefont {Lagendijk}},\ }\bibfield  {title}
  {\bibinfo {title} {Control of light transmission through opaque scattering
  media in space and time},\ }\href@noop {} {\bibfield  {journal} {\bibinfo
  {journal} {Phys. Rev. Lett.}\ }\textbf {\bibinfo {volume} {106}},\ \bibinfo
  {pages} {103901} (\bibinfo {year} {2011})}\BibitemShut {NoStop}%
\bibitem [{\citenamefont {Katz}\ \emph {et~al.}(2011)\citenamefont {Katz},
  \citenamefont {Small}, \citenamefont {Bromberg},\ and\ \citenamefont
  {Silberberg}}]{Katz2011NatPhot}%
  \BibitemOpen
  \bibfield  {author} {\bibinfo {author} {\bibfnamefont {O.}~\bibnamefont
  {Katz}}, \bibinfo {author} {\bibfnamefont {E.}~\bibnamefont {Small}},
  \bibinfo {author} {\bibfnamefont {Y.}~\bibnamefont {Bromberg}},\ and\
  \bibinfo {author} {\bibfnamefont {Y.}~\bibnamefont {Silberberg}},\ }\bibfield
   {title} {\bibinfo {title} {Focusing and compression of ultrashort pulses
  through scattering media},\ }\href@noop {} {\bibfield  {journal} {\bibinfo
  {journal} {Nat. Photon.}\ }\textbf {\bibinfo {volume} {5}},\ \bibinfo {pages}
  {372} (\bibinfo {year} {2011})}\BibitemShut {NoStop}%
\bibitem [{\citenamefont {McCabe}\ \emph {et~al.}(2011)\citenamefont {McCabe},
  \citenamefont {Tajalli}, \citenamefont {Austin}, \citenamefont {Bondareff},
  \citenamefont {Walmsley}, \citenamefont {Gigan},\ and\ \citenamefont
  {Chatel}}]{McCabe2011NatComm}%
  \BibitemOpen
  \bibfield  {author} {\bibinfo {author} {\bibfnamefont {D.~J.}\ \bibnamefont
  {McCabe}}, \bibinfo {author} {\bibfnamefont {A.}~\bibnamefont {Tajalli}},
  \bibinfo {author} {\bibfnamefont {D.~R.}\ \bibnamefont {Austin}}, \bibinfo
  {author} {\bibfnamefont {P.}~\bibnamefont {Bondareff}}, \bibinfo {author}
  {\bibfnamefont {I.~A.}\ \bibnamefont {Walmsley}}, \bibinfo {author}
  {\bibfnamefont {S.}~\bibnamefont {Gigan}},\ and\ \bibinfo {author}
  {\bibfnamefont {B.}~\bibnamefont {Chatel}},\ }\bibfield  {title} {\bibinfo
  {title} {Spatio-temporal focusing of an ultrafast pulse through a multiply
  scattering medium},\ }\href@noop {} {\bibfield  {journal} {\bibinfo
  {journal} {Nat. Commun.}\ }\textbf {\bibinfo {volume} {2}},\ \bibinfo {pages}
  {1} (\bibinfo {year} {2011})}\BibitemShut {NoStop}%
\bibitem [{\citenamefont {Mounaix}\ \emph {et~al.}(2016)\citenamefont
  {Mounaix}, \citenamefont {Andreoli}, \citenamefont {Defienne}, \citenamefont
  {Volpe}, \citenamefont {Katz}, \citenamefont {Gr{\'e}sillon},\ and\
  \citenamefont {Gigan}}]{Mounaix2016PRL}%
  \BibitemOpen
  \bibfield  {author} {\bibinfo {author} {\bibfnamefont {M.}~\bibnamefont
  {Mounaix}}, \bibinfo {author} {\bibfnamefont {D.}~\bibnamefont {Andreoli}},
  \bibinfo {author} {\bibfnamefont {H.}~\bibnamefont {Defienne}}, \bibinfo
  {author} {\bibfnamefont {G.}~\bibnamefont {Volpe}}, \bibinfo {author}
  {\bibfnamefont {O.}~\bibnamefont {Katz}}, \bibinfo {author} {\bibfnamefont
  {S.}~\bibnamefont {Gr{\'e}sillon}},\ and\ \bibinfo {author} {\bibfnamefont
  {S.}~\bibnamefont {Gigan}},\ }\bibfield  {title} {\bibinfo {title}
  {Spatiotemporal coherent control of light through a multiple scattering
  medium with the multispectral transmission matrix},\ }\href@noop {}
  {\bibfield  {journal} {\bibinfo  {journal} {Phys. Rev. Lett.}\ }\textbf
  {\bibinfo {volume} {116}},\ \bibinfo {pages} {253901} (\bibinfo {year}
  {2016})}\BibitemShut {NoStop}%
\bibitem [{\citenamefont {Pe{\~n}a}\ \emph {et~al.}(2014)\citenamefont
  {Pe{\~n}a}, \citenamefont {Girschik}, \citenamefont {Libisch}, \citenamefont
  {Rotter},\ and\ \citenamefont {Chabanov}}]{pena2014single}%
  \BibitemOpen
  \bibfield  {author} {\bibinfo {author} {\bibfnamefont {A.}~\bibnamefont
  {Pe{\~n}a}}, \bibinfo {author} {\bibfnamefont {A.}~\bibnamefont {Girschik}},
  \bibinfo {author} {\bibfnamefont {F.}~\bibnamefont {Libisch}}, \bibinfo
  {author} {\bibfnamefont {S.}~\bibnamefont {Rotter}},\ and\ \bibinfo {author}
  {\bibfnamefont {A.}~\bibnamefont {Chabanov}},\ }\bibfield  {title} {\bibinfo
  {title} {The single-channel regime of transport through random media},\
  }\href@noop {} {\bibfield  {journal} {\bibinfo  {journal} {Nat. Commun.}\
  }\textbf {\bibinfo {volume} {5}},\ \bibinfo {pages} {1} (\bibinfo {year}
  {2014})}\BibitemShut {NoStop}%
\bibitem [{\citenamefont {Shi}\ and\ \citenamefont
  {Genack}(2015)}]{shi2015dynamic}%
  \BibitemOpen
  \bibfield  {author} {\bibinfo {author} {\bibfnamefont {Z.}~\bibnamefont
  {Shi}}\ and\ \bibinfo {author} {\bibfnamefont {A.~Z.}\ \bibnamefont
  {Genack}},\ }\bibfield  {title} {\bibinfo {title} {Dynamic and spectral
  properties of transmission eigenchannels in random media},\ }\href@noop {}
  {\bibfield  {journal} {\bibinfo  {journal} {Phys. Rev. B}\ }\textbf {\bibinfo
  {volume} {92}},\ \bibinfo {pages} {184202} (\bibinfo {year}
  {2015})}\BibitemShut {NoStop}%
\bibitem [{\citenamefont {Bosch}\ \emph {et~al.}(2016)\citenamefont {Bosch},
  \citenamefont {Goorden},\ and\ \citenamefont {Mosk}}]{bosch2016frequency}%
  \BibitemOpen
  \bibfield  {author} {\bibinfo {author} {\bibfnamefont {J.}~\bibnamefont
  {Bosch}}, \bibinfo {author} {\bibfnamefont {S.~A.}\ \bibnamefont {Goorden}},\
  and\ \bibinfo {author} {\bibfnamefont {A.~P.}\ \bibnamefont {Mosk}},\
  }\bibfield  {title} {\bibinfo {title} {Frequency width of open channels in
  multiple scattering media},\ }\href@noop {} {\bibfield  {journal} {\bibinfo
  {journal} {Opt. Express}\ }\textbf {\bibinfo {volume} {24}},\ \bibinfo
  {pages} {26472} (\bibinfo {year} {2016})}\BibitemShut {NoStop}%
\bibitem [{\citenamefont {Barnes}\ \emph {et~al.}(2020)\citenamefont {Barnes},
  \citenamefont {Horsley},\ and\ \citenamefont {Vos}}]{Barnes2020JO}%
  \BibitemOpen
  \bibfield  {author} {\bibinfo {author} {\bibfnamefont {W.~L.}\ \bibnamefont
  {Barnes}}, \bibinfo {author} {\bibfnamefont {S.~A.}\ \bibnamefont
  {Horsley}},\ and\ \bibinfo {author} {\bibfnamefont {W.~L.}\ \bibnamefont
  {Vos}},\ }\bibfield  {title} {\bibinfo {title} {Classical antennas, quantum
  emitters, and densities of optical states},\ }\href@noop {} {\bibfield
  {journal} {\bibinfo  {journal} {J. Opt.}\ }\textbf {\bibinfo {volume} {22}},\
  \bibinfo {pages} {073501} (\bibinfo {year} {2020})}\BibitemShut {NoStop}%
\end{thebibliography}%

\end{document}